\documentclass[10pt]{IEEEtran}
\usepackage[nocompress]{cite}

\usepackage[hidelinks]{hyperref}
\usepackage{amsmath,amsfonts}
\usepackage{graphicx}
\usepackage[dvipsnames]{xcolor}
\usepackage[most]{tcolorbox}
\usepackage{array}
\usepackage{float}
\usepackage[caption=false,font=normalsize,labelfont=sf,textfont=sf]{subfig}
\usepackage{textcomp}
\usepackage{csquotes}
\usepackage{stfloats}
\usepackage{url}
\usepackage{booktabs}
\usepackage{tikz}
\usepackage{framed}
\usepackage{amsthm}
\usepackage{wasysym}
\usepackage{soul}
	
\newcommand{\orcid}[1]{\hspace*{0.2em}\hbox{\href{https://orcid.org/#1}{\includegraphics{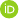}}}}
\graphicspath{{figures/}}

\newenvironment{ContributionList}{\begin{itemize}}{\end{itemize}}

\newcommand{\ra}[1]{\renewcommand{\arraystretch}{#1}}

\makeatletter
\renewenvironment{leftbar}[2]
{
	\def\FrameCommand
	{%
		{\color{#1}\vrule width 3pt}%
		\hspace{0pt}
		\fboxsep=\FrameSep\colorbox{#2}%
	}%
	\MakeFramed{\hsize\hsize\advance\hsize-\width\FrameRestore}
	\vspace*{-2mm}%
	
}
{\vspace*{-2mm}\par\unskip\endMakeFramed\vspace*{-5mm}}
\makeatother

\DeclareRobustCommand{\matrixDT}[6]{
	\begin{tikzpicture}[scale=0.075, baseline]
		\fill[#1] (0,1) rectangle ++ (1,1);
		\fill[#2] (1,1) rectangle ++ (1,1);
		\fill[#3] (2,1) rectangle ++ (1,1);
		\fill[#4] (0,0) rectangle ++ (1,1);
		\fill[#5] (1,0) rectangle ++ (1,1);
		\fill[#6] (2,0) rectangle ++ (1,1);
		\draw[darkgray] (0,0) grid (3,2); 
	\end{tikzpicture} 
}

\DeclareRobustCommand{\matrixTT}[9]{
	\begin{tikzpicture}[scale=0.05, baseline]
		\fill[#1] (0,2) rectangle ++ (1,1);
		\fill[#2] (1,2) rectangle ++ (1,1);
		\fill[#3] (2,2) rectangle ++ (1,1);
		\fill[#4] (0,1) rectangle ++ (1,1);
		\fill[#5] (1,1) rectangle ++ (1,1);
		\fill[#6] (2,1) rectangle ++ (1,1);
		\fill[#7] (0,0) rectangle ++ (1,1);
		\fill[#8] (1,0) rectangle ++ (1,1);
		\fill[#9] (2,0) rectangle ++ (1,1);
		\draw[darkgray] (0,0) grid (3,3); 
	\end{tikzpicture} 
}

\newcommand{\symYes}{\CIRCLE}
\newcommand{\symNo}{\Circle}
\newcommand{\symPartial}{\LEFTcircle}
\newcommand{\symTEAMCASEExample}{\TEAMCASEinlinesymbol{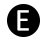}}
\newcommand{\symTEAMCASEComparison}{\TEAMCASEinlinesymbol{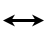}}
\newcommand{\symTEAMCASEInterview}{\TEAMCASEinlinesymbol{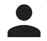}}

\definecolor{TEAMCASEColorTeaserLabelA}{RGB}{255,59,48} 
\definecolor{TEAMCASEColorTeaserLabelB}{RGB}{52,199,89} 
\definecolor{TEAMCASEColorTeaserLabelC}{RGB}{0,122,255} 

\definecolor{TEAMCASEColorTeaserLabelOne}{RGB}{255,149,0} 
\definecolor{TEAMCASEColorTeaserLabelTwo}{RGB}{179,186,218} 
\definecolor{TEAMCASEColorTeaserLabelThree}{RGB}{255,204,0} 
\definecolor{TEAMCASEColorTeaserLabelFour}{RGB}{48,176,199} 
\definecolor{TEAMCASEColorTeaserLabelFive}{RGB}{175,82,222} 
\definecolor{TEAMCASEColorTeaserLabelSix}{RGB}{162,132,94} 

\definecolor{TEAMCASEColorBGFrameworkData}{HTML}{EDF4FB}
\definecolor{TEAMCASEColorBGFrameworkModel}{HTML}{F1F7E8}
\definecolor{TEAMCASEColorBGFrameworkVisualization}{HTML}{FCE5E5}
\definecolor{TEAMCASEColorBGFrameworkKnowledge}{HTML}{FDF3E5}

\definecolor{TEAMCASEColorInfoBoxTitle}{HTML}{DF7575}
\definecolor{TEAMCASEColorInfoBoxBody}{HTML}{F8E0E0}
\definecolor{TEAMCASEColorInfoBoxNote}{HTML}{FAF0F0}

\definecolor{TEAMCASEColorCautionAdvisedBoxTitle}{HTML}{ffc712}
\definecolor{TEAMCASEColorCautionAdvisedBoxBody}{HTML}{f5e7bc}

\definecolor{TEAMCASEColorWorkflowNumber}{HTML}{647687}
\definecolor{TEAMCASEColorWorkflowBlue}{HTML}{0072C0}
\definecolor{TEAMCASEColorWorkflowRed}{HTML}{FF6666}
\definecolor{TEAMCASEColorWorkflowLila}{HTML}{A680B8}
\definecolor{TEAMCASEColorWorkflowGreen}{HTML}{97D077}
\definecolor{TEAMCASEColorWorkflowOrange}{HTML}{ff9933}

\DeclareRobustCommand\TEAMCASEcircledLetter[2]{
	\tikz[baseline=(char.base)]{
		\node[shape=circle,draw=none,fill=#1,text=white,inner sep=0.5pt] (char) {{#2}};
	}
}

\DeclareRobustCommand{\TEAMCASEinlinesymbol}[1]{%
	\begingroup\normalfont
	\raisebox{-.2\height}{\includegraphics[height=1.5\fontcharht\font`\D]{teamcase_symbols/#1}}
	\endgroup
}

\DeclareRobustCommand{\TEAMCASEDomainExpert}[2]{\mbox{\TEAMCASEhlcolor{#1}{
			#2}}}
\definecolor{TEAMCASEColorBGDomainExpertLEA}{HTML}{fff7f2}
\definecolor{TEAMCASEColorBGDomainExpertRS}{HTML}{fffdf0}
\definecolor{TEAMCASEColorBGDomainExpertSI}{HTML}{f5f0f3}
\definecolor{TEAMCASEColorBGDomainExpertLAW}{HTML}{f0f4fa}
\definecolor{TEAMCASEColorBGDomainExpertEE}{HTML}{e9f5ee}
\definecolor{TEAMCASEColorBGDomainExpertPOL}{HTML}{f2f5ed}

\DeclareRobustCommand{\DELEA}[1]{\TEAMCASEDomainExpert{TEAMCASEColorBGDomainExpertLEA}{LEA~#1}}
\DeclareRobustCommand{\DERS}[1]{\TEAMCASEDomainExpert{TEAMCASEColorBGDomainExpertRS}{RS~#1}}
\DeclareRobustCommand{\DESI}[1]{\TEAMCASEDomainExpert{TEAMCASEColorBGDomainExpertSI}{SI~#1}}
\DeclareRobustCommand{\DELAW}[1]{\TEAMCASEDomainExpert{TEAMCASEColorBGDomainExpertLAW}{LAW~#1}}
\DeclareRobustCommand{\DEEE}[1]{\TEAMCASEDomainExpert{TEAMCASEColorBGDomainExpertEE}{EE~#1}}
\DeclareRobustCommand{\DEPOL}[1]{\TEAMCASEDomainExpert{TEAMCASEColorBGDomainExpertPOL}{POL~#1}}

\newcommand\TEAMCASEsymLetter[1]{%
	\begin{tikz}[baseline=(X.base)] 
		\node (X) [color=white, fill=black, inner sep=1pt] {#1};
	\end{tikz}%
}

\DeclareRobustCommand{\TEAMCASEhlcolor}[2]{{\sethlcolor{#1}\hl{#2}}}

\newcommand{\OSFLink}{\href{https://osf.io/eap4r}{osf.io/eap4r}}

\newcommand{\papertitle}{MULTI-CASE: A Transformer-based Ethics-aware Multimodal Investigative Intelligence Framework}

\newcommand{\authorsshort}{
	{Maximilian~T.~Fischer}\orcid{0000-0001-8076-1376},
	{Yannick~Metz}\orcid{0000-0001-5955-4487},
	{Lucas~Joos}\orcid{0000-0001-7049-5203},
	{Matthias~Miller}\orcid{0000-0002-6281-2173}
	and {Daniel~A.~Keim}\orcid{0000-0001-7966-9740}
	}
\newcommand{\authorsfull}{
	\IEEEcompsocthanksitem
	M. T.\ Fischer, Y. Metz, L. Joos, M. Miller, and D. A. Keim are with the University of Konstanz. E-mail: \{max.fischer, yannick.metz, lucas.joos, matthias.miller, keim\}@uni-konstanz.de.
}

\newcommand{\keywordlist}{Intelligence analysis, communication analysis, investigative journalism, case study, ethical, evaluation, multivariate, multimodal analytics, multimedia analysis, visual analytics.}

\begin{document}

\title{\papertitle}

\author{\authorsshort
	\IEEEcompsocitemizethanks{\authorsfull}
}

\maketitle

	\begin{abstract}
		AI-driven models are increasingly deployed in operational analytics solutions, for instance, in investigative journalism or the intelligence community.
		Current approaches face two primary challenges: ethical and privacy concerns, as well as difficulties in efficiently combining heterogeneous data sources for multimodal analytics. 
		To tackle the challenge of multimodal analytics, we present \textsc{MULTI-CASE}, a holistic visual analytics framework tailored towards ethics-aware and multimodal intelligence exploration, designed in collaboration with domain experts.
		It leverages an equal joint agency between human and AI to explore and assess heterogeneous information spaces, checking and balancing automation through Visual Analytics.
		\textsc{MULTI-CASE} operates on a fully-integrated data model and features type-specific analysis with multiple linked components, including a combined search, annotated text view, and graph-based analysis.
		Parts of the underlying entity detection are based on a RoBERTa-based language model, which we tailored towards user requirements through fine-tuning and published as open-source.
		An overarching knowledge exploration graph combines all information streams, provides in-situ explanations, transparent source attribution, and facilitates effective exploration. To assess our approach, we conducted a comprehensive set of evaluations:
		We benchmarked the underlying language model on relevant Named Entity Recognition (NER) tasks, achieving state-of-the-art performance. The demonstrator was assessed according to intelligence capability assessments, while the methodology was evaluated according to ethics design guidelines.
		As a case study, we present our framework in an investigative journalism setting, supporting war crime investigations.
		Finally, we conduct a formative user evaluation with domain experts in law enforcement. Our evaluations confirm that our framework facilitates human agency and steering in security-sensitive, AI-supported analysis processes while addressing ethical and privacy concerns and providing much-needed analytical capabilities.
	\end{abstract}
	

\ifCLASSOPTIONcompsoc
\IEEEraisesectionheading{\section{Introduction}\label{sec:teamcase_introduction}}
\else
\section{Introduction}
\label{sec:teamcase_introduction}
\fi

\IEEEPARstart{A}{I-driven}
models have gained wide popularity over the last few years and have been applied successfully in numerous fields, such as natural language processing (NLP), computer vision, or predictive analytics.
Given this general trend, AI models are increasingly needed~\cite{Mitra.ModernSurveillancePrivacy.2021} and deployed in operational intelligence solutions~\cite{Bullock.AIPublicService.2020, Ganor.ArtificialHumanCounterTerror.2021}.
Corresponding application domains, such as investigative journalism~\cite{Broussard.AIJournalism.2019, Stray.AIWorkInvestigativeJournalism.2019} or the intelligence domain~\cite{Bullock.AIPublicService.2020, Hayward.AICriminologists.2021, Obendiek.PromiseSolutionismPalantirCaseStudy.2023, Mugge.SecuritizationEUTech.2023}, are particularly interesting due to their unique set of distinct challenges.
Intelligence analysts often face the task of combining numerous, heterogeneous pieces of intelligence, often tainted with uncertainty and conflicting information, forming an incomplete picture.
As discussed in previous work~\cite{Fischer.EthicalAwarenessCommAna.2022}, the first set of challenges in this regard is related to \textbf{ethical}~\cite{Shneiderman.BridgingEthicsGap.2020} and \textbf{privacy concerns}~\cite{Mitra.ModernSurveillancePrivacy.2021} due to the sensitive nature of the data and operations involved~\cite{Rigano.UsingAICriminalJustice.2019} and the high stakes in case of errors~\cite{Asaro.AIEthicsPredictivePolicing.2019, Alikhademi.PredictivePolicingFairnessAI.2022}.
Simultaneously, these domains offer opportunities for increasingly automated, tailored systems to deal with incomplete and tainted information.
This is particularly the case for \textbf{heterogeneous} and \textbf{multimodal analytics}, a second area in which existing systems often lack in functionality~\cite{Fischer.CommAID.2021,Fischer.CommunicationAnalysis.2022}.

The analysis of \textbf{individual modalities} in isolation---like network structure of the participants, named entity detection on the content, or time series analysis of the individual message intervals---often comes with limited views on the underlying information with consequences for the derived intelligence.
Not considering these aspects can reduce trust in AI systems, favor prejudices and mistakes, and also lead to legal consequences.
Further, isolated analysis requires human knowledge and intervention to semi-manually find hidden cross-matches between the modalities---a task where computational support can be highly effective, reduce domain discontinuities, and place less additional workload on the users~\cite{Fischer.CommAID.2021}.
This becomes even more important when users are no machine learning experts, thus sometimes having unrealistic expectations or misplaced trust in the systems~\cite{Fischer.EthicalAwarenessCommAna.2022, Mitra.ModernSurveillancePrivacy.2021}.
This can be the case for (business) intelligence analysts or investigative journalists, after which we modeled a case study (see Section~\ref{sec:teamcase_evaluation_case_study}).

This study is based on widespread \textbf{tasks} in intelligence, identified by the UNODC~\cite{UNODC.CriminalIntelligenceManual.2011}, which aims to answer the typical six questions: \textit{Who? What? How? Where? Why? When?}
Based on these six questions, the UNODC authors identify three common analysis tasks and methods that typically enable the answering of these questions in relevant investigations:
(1) \emph{link analysis}: searching and identifying relationships between specific entities such as persons or organizations, but also objects, locations, or events,
(2) \emph{event analysis}: correlating actions or locations alongside their timeline order,
(3) \emph{flow analysis}: understanding the connectedness as well as cause and result, for example, the flow of commodities (geolocation for physical goods or transfers of money) or the propagation of knowledge.
Other tasks described in the report involve the identification of activities, frequencies, or general data correlations.
These tasks can be primarily achieved through four main methods:
(a) keyword and semantic-based searches on text or transcripts to understand the context or find entities, (b) (social) network analysis to find connections and relations,
(c) meta-data-filters to restrict, for example, locations,
and (d) time-series analysis, for example, to identify particular communication patterns.
However, these modalities should not be considered to work in isolation but contribute individual perspectives for corroborating, enhancing, and setting each other in context.
For example, to attribute war crimes in our case study (see Section~\ref{sec:teamcase_evaluation_case_study}), our journalist Alisa leverages semantic analysis, geolocation, link-analysis, and time-correlation together with several other methods to achieve her objectives.

\begin{figure*}[t]
	\centering
	\includegraphics[width=\linewidth]{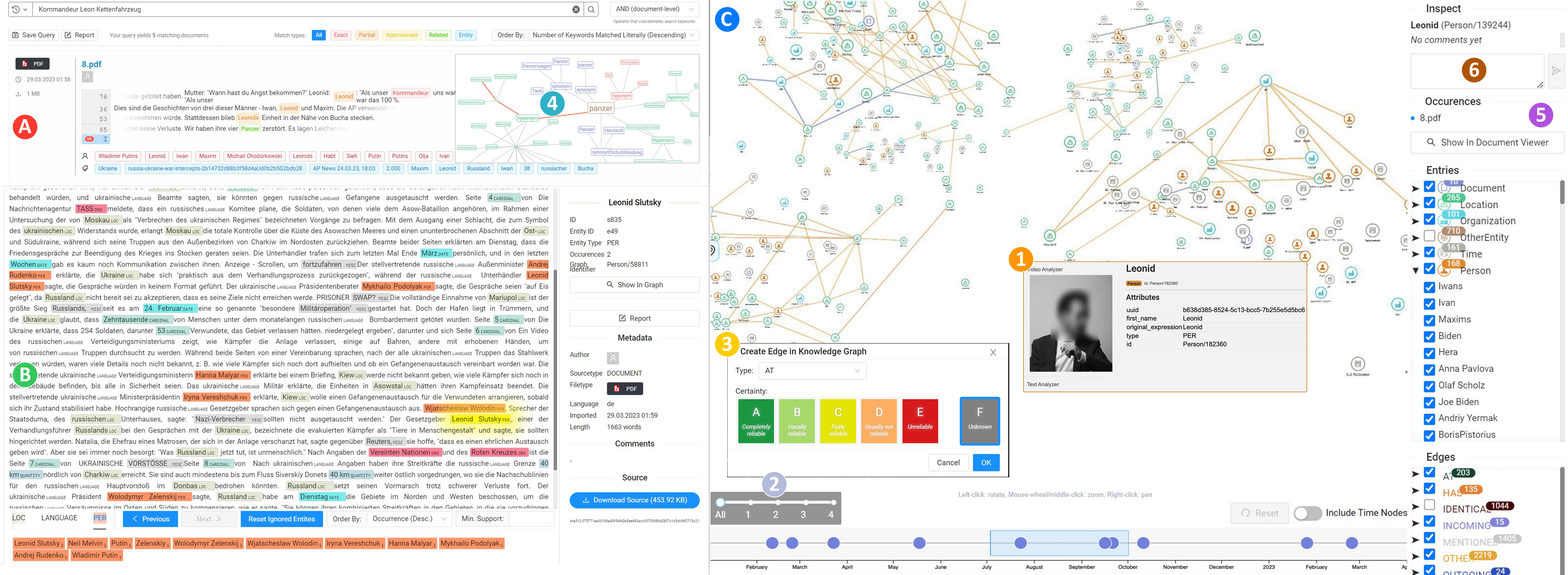}
	\caption[MULTI-CASE: A holistic visual analytics framework tailored towards ethics-aware and multimodal intelligence exploration]{%
		\textbf{MULTI-CASE: A holistic visual analytics framework tailored towards ethics-aware and multimodal intelligence exploration.}
		Built upon a fully-integrated data model, it features type-specific, graph-based analysis through individual models with multiple, linked components: \TEAMCASEcircledLetter{TEAMCASEColorTeaserLabelA}{A} a combined ontological search and result interface, \TEAMCASEcircledLetter{TEAMCASEColorTeaserLabelB}{B} an interactive textual view, and a \TEAMCASEcircledLetter{TEAMCASEColorTeaserLabelC}{C} knowledge graph interface. Additional components (not shown) include more specialized modules like video or audio analysis.
		The interface facilitates\TEAMCASEcircledLetter{TEAMCASEColorTeaserLabelOne}{1}~in-situ contextualization across modalities,\TEAMCASEcircledLetter{TEAMCASEColorTeaserLabelTwo}{2}~graph neighborhood explorations,\TEAMCASEcircledLetter{TEAMCASEColorTeaserLabelThree}{3}~relevance scoring for accountability and oversight,\TEAMCASEcircledLetter{TEAMCASEColorTeaserLabelFour}{4}~transparent source explanations,\TEAMCASEcircledLetter{TEAMCASEColorTeaserLabelFive}{5}~integrated navigation, and \TEAMCASEcircledLetter{TEAMCASEColorTeaserLabelSix}{6}~collaborative user participation.
	}
	\label{fig:teamcase_teaser_framework}
\end{figure*}

%
%
Our \textbf{objective} is to tackle the existing shortcomings in ethical and multimodal analysis for intelligence by presenting a framework for holistic communication analytics.
Many specific solutions have been proposed, but the integration and combination have received less attention.
In previous work~\cite{Fischer.CommunicationAnalysis.2022, Fischer.EthicalAwarenessCommAna.2022}, we have detailed the data and problems faced in intelligence analytics:
the need for heterogeneous data analytics capability due to the diverse set of intelligence received.
The different data types and scenario stakeholder groups like data subjects, software providers, civil society, and governmental authorities with their different branches with all their conflicting interests.
Their requirements and tasks, which we also revisit below, as well as the benefits and possible designs of visual analytics applications.

Our contribution is not intended as a fully-fledged analytics system but as an exemplary \emph{framework} for a holistic, multimodal approach to intelligence and its assessment.
Therefore, we dedicate significant time towards a comprehensive \emph{evaluation} (see Section~\ref{sec:teamcase_evaluation}), encompassing multiple perspectives, i.e., ethical aspects, capabilities, and practical considerations through use cases and expert studies.

Based on lessons learned in previous work~\cite{Fischer.CommAID.2021, Fischer.EthicalAwarenessCommAna.2022, Fischer.CommunicationAnalysis.2022}, we aim to enhance the analytical capabilities in semi-automated digital intelligence analysis, making the following \textbf{contributions}:
\begin{ContributionList}
	\item \textsc{MULTI-CASE}, an \emph{integrated} \textbf{visual exploration framework} (see Fig.~\ref{fig:teamcase_teaser_framework}) tailored towards ethics-aware \emph{multimodal} intelligence analytics in investigative journalism or criminal investigations.
	\item A RoBERTa-based NER \textbf{transformer model}, derived by fine-tuning on GottBERT~\cite{Scheible.GottBERT.2020} alongside intelligence-specific training data, which we both open-sourced at \OSFLink.
	\item An extensive \textbf{case study} showcasing \textsc{MULTI-CASE} in the context of \emph{war crime investigations} together with a \textbf{classification assessment} of its \textbf{capabilities}~\cite{Fischer.CommunicationAnalysis.2022} and \textbf{ethics design}~\cite{Fischer.EthicalAwarenessCommAna.2022}.
	\item A formative \textbf{expert evaluation} with eleven domain experts in different law-enforcement areas, validating the approach's advantages and highlighting areas for further improvement.
\end{ContributionList}

With this contribution, we fill a gap in bringing state-of-the-art performance to applications by providing an explainable visual exploration framework for multimodal intelligence analytics.
We consider our contribution primarily in the combination of existing visualization and visual analytics methodologies suitable for this domain and their detailed assessment in the context of the unique challenges faced.
Thereby, we aim to provide more insights into the often opaque workings in the intelligence domain, furthering research and a critical discussion.

\section{Related Work}
\label{sec:teamcase_related_work}

The research on \textbf{multimodal visual intelligence} analysis is sparse.
While there is significant literature on intelligence analysis in general~\cite{UNODC.CriminalIntelligenceManual.2011, Groenewald.HowAnalystsThink.2017} and some requirement studies for general intelligence analytics tools exists~\cite{Elm.DecisionSupportIntelligence.2005, Scholtz.MetricsIntelligence.2005, Fischer.CommunicationAnalysis.2022}, actual tool descriptions are rare.
If a paper evaluates an actual approach, their findings primarily focus on user acceptance while ignoring capabilities or interactive visualizations since tools are often classified and not even named publicly~\cite{Dhami.SurveyIntelligenceAnalystToolsPerception.2017}.

Research on some of the underlying techniques itself, for example, classical \textbf{Named Entity Recognition} (NER) as the foundation for comprehensive tasks like entity linking, is much more common.
Techniques evolved over time from using rule-based to more statistical systems~\cite{Nadeau.SurveyNER.2007}.
Traditionally, NER relied on annotated corpora, which posed challenges for domain transfer and new label tasks, with brittle results~\cite{Nadeau.SurveyNER.2007, Li.SurveyNER.2020}.
However, with the advent of deep learning-based approaches, such as BERT~\cite{Devlin.BERT.2018}, the landscape has changed, and transfer learning (i.e., adapting pre-trained models to shorten training times for new tasks) can cope with much smaller amounts of annotated text.
This has significantly improved the adaptability and efficiency across various domains and tasks, making knowledge transfer and few-shot labeling easier~\cite{Devlin.BERT.2018, Lee.TransferLearningNER.2017}, which can be leveraged in investigative tools.

Similarly, advances in \textbf{ethical design}~\cite{Barbosa.WhatsappEthics.2019, Fischer.EthicalAwarenessCommAna.2022}, like the concept of providing guidance~\cite{Sperrle.Lotse.2022}, visualizing hidden uncertainties~\cite{Zytek.Sibyl.2022}, or ensuring provenance~\cite{Correll.EthicalDimensionsVis.2019} as well as \textbf{ privacy considerations}~\cite{Fischer.EthicalAwarenessCommAna.2022}, like selective masking~\cite{Tu.TowardMassVideoSurveillance.2021}, federated learning~\cite{Geyer.FederatedLearning.2017}, or data perturbation~\cite{Shynu.FuzzyDataPerturbationPrivacy.2020} have been made.
Also, \textbf{insular solutions} like Pajek~\cite{Batagelj.Pajek.2002} for social network analysis, Maltego~\cite{Schwarz.DesignMaltego.2021} or InSight2~\cite{Kodituwakku.InSight2.2020} for link analysis, or Cosmos~\cite{Dowling.InteractiveSensemakingText.2019} for semantic text analysis exist but do not combine modalities.

Within the visualization community, \textbf{multimodal multimedia analysis}~\cite{ZahalkaWorring.MultimediaAnalytics.2014} can be considered partly related:
Several approaches have been proposed to consider different aspects of multimedia content simultaneously, like the presentation styles and techniques~\cite{Wu.MultimodalVideoCollections.2020}, the emotional coherence~\cite{Zeng.EmoCo.2020}, or the automation of explicit content through video moderation~\cite{Tang.VideoModerator.2022}.
While these approaches propose valuable insights into how (primarily visual) media can be analyzed and set into context, many of the approaches target very specific applications, and very few in this domain truly support a holistic approach to analyzing \emph{generic} pieces of intelligence, which also includes text-based information.
Further, Zahalka and Worring presented a pathway to comprehensive multimedia analytics, detailing a general four-tiered multimedia analytics model and discussing it alongside how it may support addressing the semantic and pragmatic gap encountered in actual systems~\cite{ZahalkaWorring.MultimediaAnalytics.2014}.
This follows a similar overall direction as our research, however, with one particular difference:
The model is applicable in general for the analysis of multimedia data and also with a particular focus on such data, for example multimedia collections of images.
While some aspects overlap, these collections of images do not necessarily have a underlying storyline, may come from any collection mechanism (e.g., underwater camera), an the model primarily focuses on a multimodal analysis of multimedia with additional metadata (e.g., annotated text or features).
Our approach instead focuses primarily on communication between humans, emphasizing much more the interactive aspects of the information exchange via various modalities over time.

The research on leveraging \textbf{visual analytics for intelligence applications}~\cite{DeckerStasko.VAExploration.2009, Kang.EvaluatingVAInvestigative.2009, Kang.CharacterizingIntelligenceAnalysis.2011, Lu.IntegratingPredictiveAnalytics.2014} had its prime in the mid-to-late 2000s, with frameworks such as VIM~\cite{Keahey.VIMIntelligenceAnalysis.2004} or Jigsaw~\cite{Stasko.Jigsaw.2007}.
Both primarily focus on text documents (and not so much multimedia), and only a few approaches~\cite{Fischer.CommAID.2021} were proposed later on.
Therefore, this area seems to be one of those few domains where commercial research has outpaced academic, scientific research for now.

In the context of actual usage---also for commercial systems---we surveyed related communication analysis systems~\cite{Fischer.CommunicationAnalysis.2022}, where we identified four publicly known intelligence systems in wider use:
DataWalk~\cite{DataWalk.2020} and Nuix Discover / Investigate~\cite{Nuix.DiscoverInvestigate.2020} are sometimes used, while the market leaders are IBM i2 Analyst's Notebook~\cite{IBM.AnalystsNotebook} along with Palantir Gotham / Foundry / Meta-Constellation~\cite{Palantir.Gotham.2020}.
While they cater to government applications, parts are commercially available and are used by international banks, advertisers, manufacturers, telecommunication providers, media organizations, and NGOs~\cite{Palantir.Gotham.2020}.

To our knowledge, no new visual analytics approaches to intelligence have been publicly proposed since our recent survey of AI-driven intelligence applications~\cite{Fischer.CommunicationAnalysis.2022}, also available as an \textbf{interactive browser} at \href{https://communication-analysis.dbvis.de}{https://communication-analysis.dbvis.de}.
Regarding practical usage, the ongoing shift from IBM i2 to Palantir seems to accelerate. %
Palantir's solutions (in particular Meta-Constellation) are also employed effectively~\cite{Scott.UkrainePalantir.2022} by Ukraine in its defense against Russia in coordinating their military.

The \textbf{academic research} on this topic has been falling short, with problematic consequences for accountability and oversight, which has also been realized by some key stakeholders.
For example, in the European Unions Horizon 2020 funding period alone, projects such as ASGARD (\href{https://cordis.europa.eu/project/id/700381}{700381}), MAGNETO (\href{https://cordis.europa.eu/project/id/786629}{786629}), STARLIGHT (\href{https://cordis.europa.eu/project/id/101021797}{101021797}), COPKIT (\href{https://cordis.europa.eu/project/id/786687}{786687}), and AIDA (\href{https://cordis.europa.eu/project/id/883596}{883596}) (some still ongoing) have been funded, although preliminary results show insular capabilities.
For the upcoming Horizon Europe funding period, several calls have been proposed (e.g., HORIZON-CL3-2023-FCT-01).
With slight deviations, they all aim to increase analytical big data capabilities for law enforcement.
In the US, similar research is often conducted by national laboratories but mostly remains classified.

While many visualization approaches can be leveraged for intelligence, only few consider the combination of challenges faced in this particular domain, including the inherent uncertainty and inter-modality, while even fewer evaluate them consistently and publish the results, which is the goal of this work.

\section{Methodology: Model Development}
\label{sec:teamcase_methodology}
In previous work~\cite{Fischer.CommAID.2021}, we have presented a matrix-based, holistic communication analysis framework through semantic zooming.
As our studies have shown, however, despite the potential benefits in scalability, matrices are uncommon for many analysts, which are used to graph- and relationship-based visualizations.
Further, semantic zooming is space-limited in the amount of context information in the upper layers.
We, therefore, aim to explore an \emph{orthogonal design}, with \emph{two key advancements}:
(a) Following a similar modular approach, we leverage a more powerful \textbf{fully-integrated data model} (structuring and relating the intelligence information pieces) that also supports multimodality.
(b) Instead of matrix-based semantic zooming, we use a \textbf{graph-based overview} with several linked views and integrated specialized views.

This decision is based on the task descriptions and requirements described in the UNODC report~\cite{UNODC.CriminalIntelligenceManual.2011} described above, as well as feedback from several domain experts in law enforcement, which state the following three \textbf{user requirements} for such a framework:
(1) A centralized, multimodal platform for collaborative case working.
(2) Assistance in labor-intensive tasks such as big data analytics.
(3) Transparency and reliability.

This reflects their need to \emph{work collaboratively} on a case together with their colleagues on larger investigations, needing to share results or to leverage knowledge generated by colleagues investigating specific aspects of a case by collaboratively working on a shared data space and being able to access the information in-situ.
Due to the sheer volume and sometimes repetitive tasks, support by automation and AI is considered essential while being reliable and understandable.
All the while, the analysis steps taken need to be transparent and reproducible for accountability.
Guided by these overall principles, we further justify individual design decisions and capabilities while describing the system design in Section~\ref{sec:teamcase_system}.

\begin{figure*}[t]
	\includegraphics[width=\linewidth]{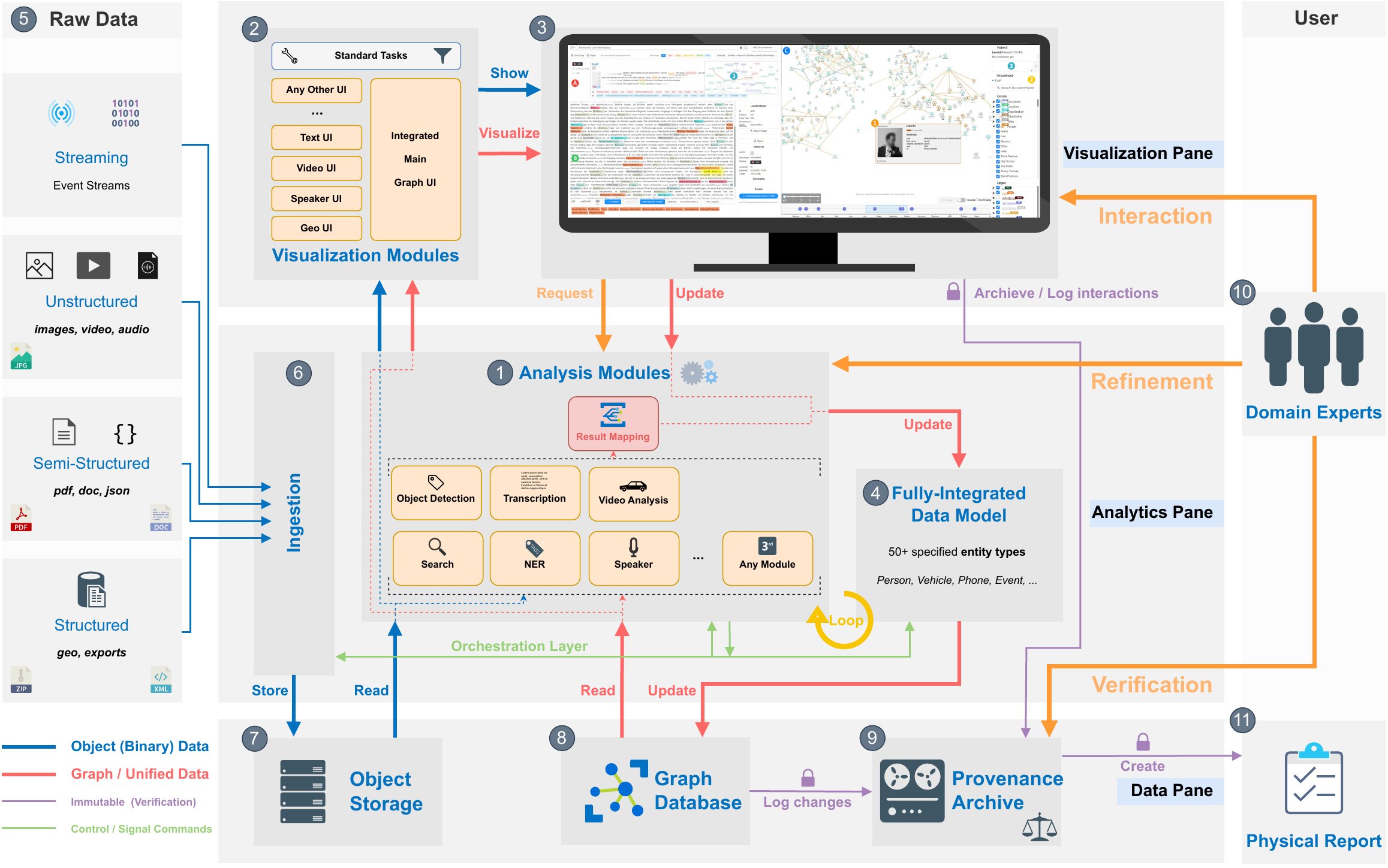}
	\caption[High-level architecture of MULTI-CASE]{\textbf{High-level architecture of \textsc{MULTI-CASE}}, highlighting the main components \TEAMCASEcircledLetter{TEAMCASEColorWorkflowNumber}{1}-\TEAMCASEcircledLetter{TEAMCASEColorWorkflowNumber}{9}, like the \TEAMCASEcircledLetter{TEAMCASEColorWorkflowNumber}{1}~analysis modules or the \TEAMCASEcircledLetter{TEAMCASEColorWorkflowNumber}{3}~graph UI with \TEAMCASEcircledLetter{TEAMCASEColorWorkflowNumber}{2}~visualization modules, the \TEAMCASEcircledLetter{TEAMCASEColorWorkflowNumber}{4}~fully-integrated graph data model, as well as different data paths and user handling, like the \textcolor{TEAMCASEColorWorkflowOrange}{---} analytics workflow.
		A detailed description is provided in Section~\ref{sec:teamcase_system}.
	}
	\label{fig:teamcase_architecture}
\end{figure*}

One central aspect of intelligence analytics is the analysis of communication~\cite{Fischer.CommAID.2021}.
However, common international NER labeling schemes (e.g., PER, ORG, LOC, OTH) often do not meet the specialized requirements for investigations since they are too ambiguous and not specialized enough, requiring more narrow tag categories~\cite{Mohr.GraphingNSS.2013, Ratinov.DesignChallengesNER.2009}.
In practice, specialized \textbf{NER model development} for semantic understanding is still challenging, with many pitfalls, although the attention-based transformer architecture~\cite{Vaswani.AttentionIsAllYouNeed.2017} has significantly increased the accuracy compared to previous neural models.
Therefore, as part of this work, we track the necessary steps for training and deploying transformer models, including interactive tools for labeling, while also highlighting major lessons learned.
The necessary steps range from choosing a suitable base model, preparing representative training data, then training and evaluating, to finally supervising and validating the model in deployment and adapting it in the face of changing language patterns, terms, or requirements.
As a result, we provide a strong baseline NER transformer model with a large set of relevant entity labels to simplify future applications.
For the underlying language model, we considered existing models from the Huggingface transformers~\cite{Wolf.HuggingFaceTransformers.2019} library based on evaluation performance on the GermEval14 dataset~\cite{Benikova.Germeval.2014}, a well-known dataset for German NER recognition.
For German natural language processing, we considered two language models: The RoBERTa-based GottBERT~\cite{Scheible.GottBERT.2020} and \textit{BERT-base-german-cased}~\cite{Schwimmbeck.BertHugginface.2022} based on the original BERT transformer architecture~\cite{Devlin.BERT.2018}.
Additionally, we chose a strong multi-lingual baseline (XLM-RoBERTa)~\cite{Conneau.UnsupervisedCrossLingualRepr.2019}.

In general, the creation of specific training datasets, for example, through \textbf{labeling} of domain-specific datasets, is often tedious and error-prone.
Therefore, we implemented an interactive labeling tool that is compatible with the \textsc{MULTI-CASE} framework, allowing us to label and subsequently review a given document collection on a large scale, facilitating the easy creation of ground truth training datasets in specific domains, like intelligence.
This is particularly relevant in our application because it utilizes a large set of custom-named entity labels for domain-specific analysis.
Many non-English models only provide standard categories like \textit{PERSON}, \textit{LOCATION}, \textit{ORGANIZATION}, and \textit{MISC}.
However, based on expert feedback, custom categories like \textit{EVENT} or \textit{PRODUCT} and more fine-grained time and numeric labels were introduced, with the full list shown in Table~\ref{tab:teamcase_model_ner_accuracy}.
We provide an enhanced, re-tagged version of GermanNER alongside our model at \OSFLink.


For the \textbf{training}, we apply a train/validation/test split of 70/15/15 of the full mixed dataset (domain-specific and re-tagged corpus data).
We train each baseline model with Adam~\cite{Kingma.AdamOptimizer.2014}, weight decay~\cite{Loshchilov.DecoupledWeightDeay.2017}, and 0.1 dropout.
We also experimented with a slanted triangular learning rate (i.e., using a warm-up and linearly decaying learning rate)~\cite{Howard.FineTunedLangModelForClassification.2018} and found a slight positive effect on final performance.
Early stopping was implemented based on the validation F-score with a number of patience steps of 10.
Fine-tuning of all models was performed on a single RTX4000 GPU.
We report the full set of hyperparameters and additional results at \OSFLink.

After training, we evaluate the model performance alongside other base models on a held-out test set and describe the results in Section~\ref{sec:teamcase_evaluation_model}.
We also note the recent advances by Large Language Models (LLMs), which can drastically improve specialized NER-tagging through zero- or few-shot learning, in the outlook in Section~\ref{sec:teamcase_limitations_future_work}.

\section{System Design}
\label{sec:teamcase_system}
The proposed architecture for our framework and the fully-integrated graph data model described in the following is shown in Fig.~\ref{fig:teamcase_architecture}.
When necessary, we also detail the expert reasoning and the ethical considerations behind individual design decisions while also referring to Sections~\ref{sec:teamcase_methodology} and~\ref{sec:teamcase_evaluation_expert_study} as well as~\ref{sec:teamcase_evaluation_ethics_design} for further discussions on these topics.
The guideline numbers for the ethical and privacy reasoning (e.g., C1-6, R1-5, A1-6) refer to the nomenclature established in previous work~\cite{Fischer.EthicalAwarenessCommAna.2022}.

Overall, the system consists of individual plugins \TEAMCASEcircledLetter{TEAMCASEColorWorkflowNumber}{1}~{\color{TEAMCASEColorWorkflowBlue}Analysis Modules} for specific analysis tasks and data types, a \TEAMCASEcircledLetter{TEAMCASEColorWorkflowNumber}{3}~{\color{TEAMCASEColorWorkflowBlue}Main Graph-based UI} together with specialized \TEAMCASEcircledLetter{TEAMCASEColorWorkflowNumber}{2}~{\color{TEAMCASEColorWorkflowBlue}Visualization Modules} (e.g., text analysis or video-analysis) for a web-based exploration.
This fulfills the demand by experts to be capable of specialized analysis that interfaces with an overall case working framework.
Similarly, the heavy computations are run on a centralized server, while the interface nowadays is a standard web-based approach running on a regular (or thin) client.
One key aspect of the overall system is the \TEAMCASEcircledLetter{TEAMCASEColorWorkflowNumber}{4}~{\color{TEAMCASEColorWorkflowBlue}Fully-Integrated Data Model} stored in a \TEAMCASEcircledLetter{TEAMCASEColorWorkflowNumber}{8}~{\color{TEAMCASEColorWorkflowBlue}Graph Database}, which acts both as a conceptual abstraction layer between modules and a central source of shared knowledge.
This enables the experts to work on a consistent data set in an integrated environment and not lose information compared to switching between applications, increasing Efficiency (A4) while addressing the working together of machines and users (C5).
Supporting roles fall to the \TEAMCASEcircledLetter{TEAMCASEColorWorkflowNumber}{7}~{\color{TEAMCASEColorWorkflowBlue}Object Storage} to store any input and intermediate data and the \TEAMCASEcircledLetter{TEAMCASEColorWorkflowNumber}{9}~{\color{TEAMCASEColorWorkflowBlue}Provenance Archive} as a {\color{TEAMCASEColorWorkflowLila}revision-safe storage}, which is considered essential for Opacity (C3) and Accountability (C6).
The\TEAMCASEcircledLetter{TEAMCASEColorWorkflowNumber}{10}~{\color{TEAMCASEColorWorkflowBlue} domain experts} can communicate with the system by interacting with the visualization, forming a collaborative Human-Machine-Configuration (C5), refine the display through analysis parameters, as well as {\color{TEAMCASEColorWorkflowLila}verify} the results, which increases understanding and fosters trust and works against Lack of Accountability (R1), while enabling Human Oversight (R5) and also facilitating a critical reflection (R4).
This verification is available both in the interface and in a \TEAMCASEcircledLetter{TEAMCASEColorWorkflowNumber}{11}~{\color{TEAMCASEColorWorkflowBlue}physical report}, which the experts still need to document their findings in a structured way.

\subsection{Data Model}
Diverse types of \TEAMCASEcircledLetter{TEAMCASEColorWorkflowNumber}{5}~{\color{TEAMCASEColorWorkflowBlue}Raw Data} are supported, ranging from unstructured data (images, video, audio), over semi-structured documents (e.g., PDF documents), to structured data types (like geolocation tracks or exports), as well as streaming data.
The needs of the domain experts naturally vary here depending on their organization and tasks, but typically the first two types are the most common ones.
The input is only limited by the plugin analysis modules.
When {\color{TEAMCASEColorWorkflowBlue}data is} \TEAMCASEcircledLetter{TEAMCASEColorWorkflowNumber}{6}~{\color{TEAMCASEColorWorkflowBlue}ingested}, it is stored in the \TEAMCASEcircledLetter{TEAMCASEColorWorkflowNumber}{7}~{\color{TEAMCASEColorWorkflowBlue}Object Storage}.
Based on the input type, the {\color{TEAMCASEColorWorkflowGreen} Orchestration Layer} selects one or more analysis modules for knowledge extraction, for example, NER for text documents.
The main results are mapped to the \TEAMCASEcircledLetter{TEAMCASEColorWorkflowNumber}{4}~{\color{TEAMCASEColorWorkflowBlue}Fully-Integrated Data Model} stored in the \TEAMCASEcircledLetter{TEAMCASEColorWorkflowNumber}{8}~{\color{TEAMCASEColorWorkflowBlue}Graph Database}.
For example, for NER, this could be the detected \emph{entities}, like persons, location, or dates, as well as their \emph{relations}, while for video analysis, an object like a car along its properties and a relationship to time and location.
Two aspects are of primary importance:

\textbf{(1)} the \textbf{data model} ideally has to be as mutually exclusive and collectively exhaustive as possible.
The data model was designed with several domain experts and generalized from existing case models like IMP (Information Model Police).
In our case, we arrived at 50+ hierarchical \emph{entities} (graph nodes) and 10+ \emph{relationship} types (graph edges), trying to find the right balance between a generic data model and enough specialization.
While a very generic data model allows for the reflection of virtually all analysis results, the automatic conclusions, connections, and information enrichment in such a case can remain very limited.
In contrast, a highly specialized data model allows to reflect on the findings with high precision and enables many automated conclusions. However, it always poses the danger of being too specified (i.e., available properties on a type) to capture all relevant information.
Indeed, the principle design is flexible, subject to change, and can be adapted by adding more specialized entities or data fields.
Analysis modules are change-agnostic if the entities and attributes they work with are untouched.
In our case, we derived everything from a root \emph{Thing}, with \emph{Entity}, \emph{Event}, \emph{Datetime}, \emph{Location}, and \emph{Document} as the first hierarchical layer, each having further subtypes (e.g., \emph{Person} or \emph{PhoneCall}).
For example, a \emph{Timespan}, as a subtype of \emph{Datetime}, represents a specific time range and can be related to a \emph{PhoneCall} via a relationship, which in turn may be related to specific phone numbers, which again might be related as belonging to actual persons.
Attributes for each entity store associated information.
Through \emph{relationships}, one can also model source attribution (source document and analysis module) and\TEAMCASEcircledLetter{TEAMCASEColorTeaserLabelThree}{3}confidence scores, e.g., based on the 6x6 intelligence scoring~\cite{UNODC.CriminalIntelligenceManual.2011}, which many analysts are well familiar with, strengthening Literacy (A5).
This scoring can have an influence on automatic decision-making: when certainties are considered by algorithms, this can support working towards Preventing Automated Inequality (R3) and limit Exaggerated Expectations (C4) and Discriminatory bias (C1) through manual priming.
Simultaneously, the opposite could also be true, where the system warns a user of inherent prejudice evident in analysis choices.

\textbf{(2)} The data model allows a structured information \textbf{exchange} and also \textbf{information enrichment process} between modules, which the experts consider essential.
{\color{TEAMCASEColorWorkflowRed}Updates of the data model} can trigger subsequent runs of other analysis modules when they have signed up for specific creations/updates:
for example, an imported audio file might be analyzed first by a speaker detection (with the creation of a specific audio entity), then by a speech-to-text transcription (with a text entity), and then by a NER process, which can result in an {\color{TEAMCASEColorWorkflowRed}enrichment of the graph} with the conversation content through multiple entities (e.g., persons, location, or times).
All changes (creations, updates, hidings) in the graph data are {\color{TEAMCASEColorWorkflowLila}logged via a write once}, read many \TEAMCASEcircledLetter{TEAMCASEColorWorkflowNumber}{9}~{\color{TEAMCASEColorWorkflowBlue}Provenance Archive}.

\subsection{Component Integration}
The individual \TEAMCASEcircledLetter{TEAMCASEColorWorkflowNumber}{1}~{\color{TEAMCASEColorWorkflowBlue}Analytics Modules} like NER or transcription are designed as plugins and can be flexibly combined depending on the analytical needs, allowing for Customization (A6) and ensuring User Agency (A1) of the experts.
In this work, we primarily focus on the search and NER modules as an exemplary prototype developed by us, while other modules are provided as open source (e.g., transcription via Whisper~\cite{Radford.Whisper.2022}) or by commercial partners.
During startup, the modules register themselves, their supported data types for ingestion, and the graph change listeners via the {\color{TEAMCASEColorWorkflowGreen} Orchestration Layer}.
Further, each analytics module can register custom \emph{context actions} (e.g., show similar persons) and \emph{preview handlers} (e.g., picture or video player), which are integrated into the \TEAMCASEcircledLetter{TEAMCASEColorWorkflowNumber}{3}~{\color{TEAMCASEColorWorkflowBlue} Main Graph UI}, allowing for a tight coupling between the UI and individual modules functions in \TEAMCASEcircledLetter{TEAMCASEColorWorkflowNumber}{2}~{\color{TEAMCASEColorWorkflowBlue}specialized UIs}, supporting the mental mapping of the experts.

\subsection{Interfaces and Interaction Principles}
The interfaces are web-based, and the provided views are \textbf{tightly coupled} and \textbf{inter-linked}, strengthening the Human-Machine-Configuration (C5) and the User Agency (A1) through Opacity (C3).
Entities are consistently mapped via the unified, fully-integrated data model, allowing for the enrichment of information within the main graph-based overview and across views.

The main interface to start explorations is the \TEAMCASEcircledLetter{TEAMCASEColorWorkflowNumber}{3}~{\color{TEAMCASEColorWorkflowBlue} Main Graph UI} (see also \TEAMCASEcircledLetter{TEAMCASEColorTeaserLabelC}{C} in Fig.~\ref{fig:teamcase_teaser_framework}).
It provides a highly scalable \emph{GPU-based rendering} of a \emph{Knowledge Graph} (a network-based visualization of the interconnected data items and their relationships), together with several linked views.
This graph-based overview is less scalable than a matrix-based approach~\cite{Fischer.CommAID.2021}, however, aligns more closely with the mental image of analysts when exploring a network, as link charts have been used in investigative work for a long time~\cite{Sparrow.NetworkAnalysisCriminalIntelligence.1991}.
The user can navigate this graph with a mouse and keyboard, select, hover, move, and (context) click individual nodes (data or extracted information items) and edges (their relations).
The graph uses a selectable 2D or 3D node-link representation and is rendered using a force-directed layout.
Strengths are calculated using centralizing, link, and charge forces based on a Barnes–Hut approximation.
While this graph is initially automatically generated, the expert can (and is expected to) explore, interact, add, modify, and enhance it while working on the case.
When modifying or judging information, user confidence in relationships (edges) can be encoded using the 6x6 intelligence scoring system for \TEAMCASEcircledLetter{TEAMCASEColorTeaserLabelThree}{3}relevance grading.
The default confidence is F (Unknown), and for automated decisions that have not been manually reviewed, never above C (fairly reliable) to prevent Automated Inequality (R3) and wrong conclusions.
All the interactions happen within the graph view or via individual visualization modules, which are reachable via the registered context actions and context menus, allowing for seamless transitions, which are appreciated by users.
\begin{figure}
	\centering
	\includegraphics[width=\linewidth]{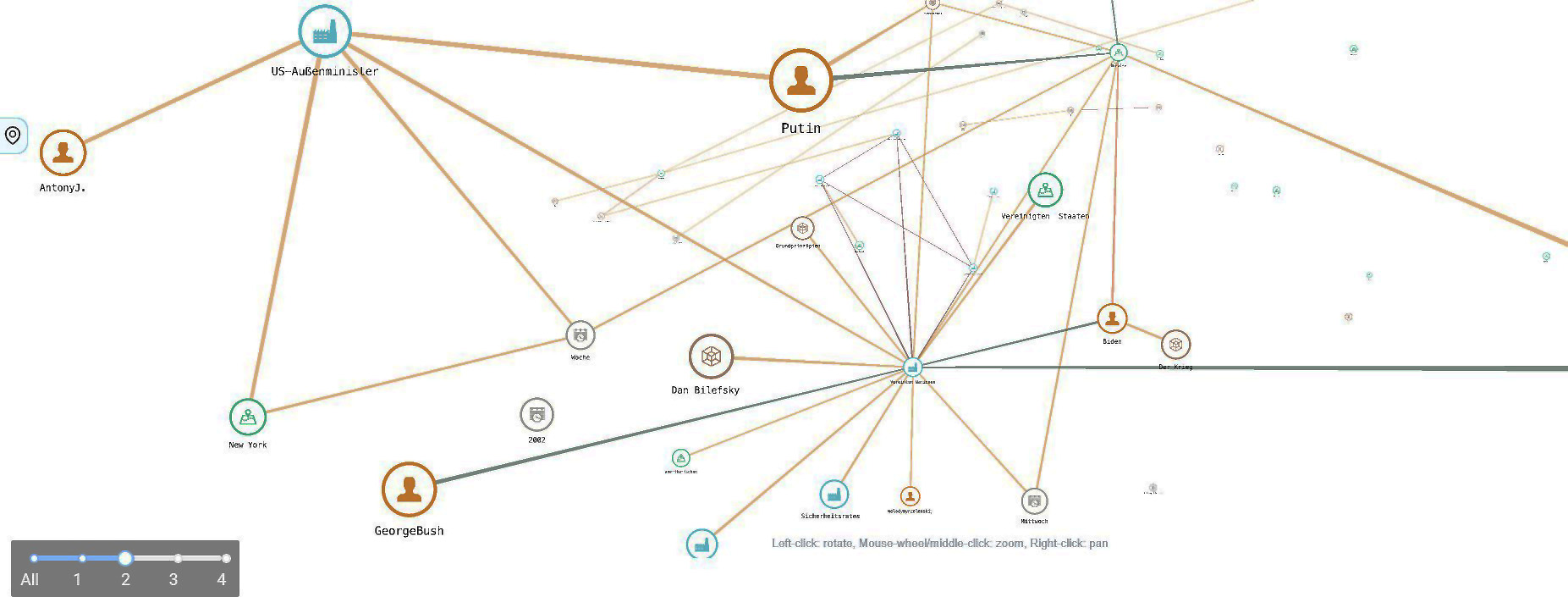}
	\caption[Neighborhood exploration]{%
		\textbf{Neighborhood exploration} \TEAMCASEcircledLetter{TEAMCASEColorTeaserLabelTwo}{2}, acting as a magnifying spotlight to show a manageable local context for a seamless exploration.
	}
	\label{fig:teamcase_neighborhood_exploration}
\end{figure}

The visual interface to the graph model has several features to enable Customization (A6) and User Agency (A1):
A \emph{sidebar} on the right offers several features: (1) control options for visualization (e.g., color, line thickness), layouting (force-direction layout strengths), and modes (e.g., 2D/3D-dimensionality, display modes for exploration like only displaying cross-matches, i.e., results from multiple documents) allow for customization and task-specific adaption,
(2) an interactive \emph{search} functionality allows filtering the graph quickly, (3) a \emph{context display} shows information about a selected entity, and---leveraging the integrated graph model---occurrences, e.g., in text documents,
(4) an overview of all available nodes and edges grouped by types, (de-)selectable individually or in groups.

A \emph{timeline} at the bottom shows both datetime information as part of the Knowledge Graph and document times, allowing for brushing and filtering to optimize the graph view and empower investigators to follow an event- and time-based workflow in alignment with their exploration.
When hovering over documents, only these are shown, while selecting zoomable and shiftable ranges restricts the shown parts of the graph.
As can be seen from the examples, the amount of information displayed in the graph view is typically quite large, which hampers exploration.
Therefore, a\TEAMCASEcircledLetter{TEAMCASEColorTeaserLabelTwo}{2}\emph{Neighborhood Exploration}, acting similar to a magnifying glass or spotlight, allows to show the local neighborhood of a node (for example, 3 or 4 steps), and clicking any visible node transitions to the new neighborhood, allowing for a seamless exploration with a manageable amount of local, contextual information displayed without overloading the users, which can improving Efficiency (A4).
Another approach to reducing the amount of clutter is to selectively merge confirmed relations to clusters, for example, aliases for persons or create groups.
A slider allows for a confidence level based on the 6x6 system, which means that automated decisions without manual verification are never categorized as (very) likely (A or B), preventing  Automated Inequality (R3) and enforcing Fairness (A3) and critical reflection (R4) through Human Oversight (R5). 

Due to the amount of information shown (for typical investigations, this can be 30k nodes and 100k edges), we need to use several techniques to achieve 60+ fps performance:
The graph is rendered entirely on the GPU and leverages instancing and custom shaders.
This results in, once set up, a fixed-sized geometry of a few hundredth vertices and three WebGL draw calls (nodes, edges, labels), resulting in efficient rendering performance.
Much of the visualization and visibility status is controlled from within the shaders, with crafted texture atlases and mipmapping for efficient textures, especially for nodes and text labels.
To render more than 0.5 million text characters in real-time, we use a pre-generated font texture atlas and supply each node label instance with its correct, fixed-size ASCII-Code label (Unicode would be possible, but increase the texture size).
This supply of instance-specific data (e.g., labels, position, node render state) is achieved through uniform buffer objects, acting similarly to a memory map, which is highly efficient.
The sidebar uses virtual lists to render on demand, further reducing DOM usage.
However, the number of nodes and edges is still limited by JavaScript and Browser performance.

The NER module offers an \TEAMCASEcircledLetter{TEAMCASEColorTeaserLabelA}{A} ontological search and \TEAMCASEcircledLetter{TEAMCASEColorTeaserLabelB}{B} textual view (see Fig.~\ref{fig:teamcase_teaser_framework}) as UI components.
In the UI, a \TEAMCASEcircledLetter{TEAMCASEColorTeaserLabelOne}{1}\textbf{context overlay} can be shown, for example, over a person's name with a preview image of a person together with other meta-data.
This reduces domain boundaries and relieves the mental load of the users.
Text understanding can be helped (see Section~\ref{sec:teamcase_evaluation_expert_study}) by color-coding named entities according to type and offering aggregation and interactions.
Linked views at the bottom show all entities in the document grouped by type and ordering, e.g., by count, can be used to quickly navigate between occurrences through auto-scrolling, highlighting, and stepping.

The \textbf{ontological search} uses multiple (de-) selectable semantic search modes (exact match, substring match, fuzzy match, or ontological match).
The latter allows searching \emph{semantically} instead of guessing the correct keywords.
This ontological search is considered very beneficial by the experts, as it reduces the burden on them to know the exact terms used but more generically describes the concept of what they are looking for.
Search results are shown with specific probability scoring based on the distance (steps) taken in an ontology database, linking different properties.
One example would be to search for "accommodation" and get results with "hut", "hotel", or "cottage".
The quality of the results, of course, depends on the extensiveness of the ontology, which often has to be adapted domain-specifically.
Here, the experts can modify the ontology \emph{on the fly}, e.g., to adapt to specific codewords.

Another type of interaction resulting from the tight integration comes even closer to the {\color{TEAMCASEColorWorkflowOrange}traditional visual analytics loop}:
While updating analysis parameters within a module usually only affects this module's results, through the fully-integrated data model and module listeners, it becomes possible to achieve \emph{inter}-module exploration and refinement, coming closer to the expected levels of automation by current users.
For example, when several speakers in audio files are recognized, and the transcripts are polluted by some of the speakers being background noise, the user can manually deselect the speakers, resynthesizing the audio, and the downstream analysis is automatically re-run, i.e., transcription and then knowledge extraction through NER.
Old results can, in this process, be hidden (i.e., flagging the old document and its inference) to avoid an over-cluttering of the graph, which is considered extremely relevant by the experts to allow them to focus on relevant information only but can also be used to preserve Privacy (A2, C2).

While users work with the application, all performed actions are {\color{TEAMCASEColorWorkflowLila}logged to achieve provenance}, provide Accountability (C6), as well as prevent abuse through Human Oversight (R5).
\section{Evaluation}
\label{sec:teamcase_evaluation}
We conducted a thorough evaluation of our approach, including feedback from multiple perspectives, to determine the effectiveness of the system.
To showcase the practical usefulness of our approach, we present a case study in an investigative journalism setting, supporting war crime investigations (see Section~\ref{sec:teamcase_evaluation_case_study}).
To scrutinize the ethical and privacy risks involved, we then evaluate our approach based on ethics design guidelines~\cite{Fischer.EthicalAwarenessCommAna.2022} for intelligence applications (see Section~\ref{sec:teamcase_evaluation_ethics_design}).
To judge the resulting capabilities of the developed framework, we use a state-of-the-art intelligence capability assessment~\cite{Fischer.CommunicationAnalysis.2022} (see Section~\ref{sec:teamcase_evaluation_intelligence_capabilities}).
To assess the quality of the underlying language model, we performed benchmarks on relevant NER-task, achieving state-of-the-art performance (see Section~\ref{sec:teamcase_evaluation_model}).
Finally, to evaluate the system from an expert perspective, we conducted a formative user evaluation with eleven domain experts in law enforcement (see Section~\ref{sec:teamcase_evaluation_expert_study}).

\subsection{Case Study}
\label{sec:teamcase_evaluation_case_study}
In the following, we describe a simplified, \emph{artificial} case study modeled after real-world workflows seen in \textbf{investigative journalism}.
Here, we describe the process of identifying, placing, attributing, and documenting \textbf{war crimes}.
We have chosen this example due to its high relevance, the high analysis stakes both for the victims as well as innocent persons, and the plausible availability of large amounts of multimodal data.

\textbf{Goal} --- Alisa is an aspiring journalist for the respected newspaper \textit{The Custodian}.
She has been reporting about a brutal war in her home country for months now.
While there have been some high-profile reports on war atrocities, she knows this is just the tip of the iceberg, and many people are missing. 
After reading some OSINT (Open Source Intelligence) reports, she wonders if she can also find out more about the forgotten victims of war.
Simultaneously, she wants to see the perpetrators held accountable, so she aims to document her findings meticulously and hand her chain of evidence over to the ICC (International Criminal Court), which has started pre-trial investigations.

\textbf{Data Collection} --- She starts off by collecting raw data:
From various online sources reporting about the war, like Telegram, she exports messages, images, audio, and videos.
From a friend and contact working for a large telecommunication provider, she gets a large dump of telephone calls and texts originating from foreign cell phone numbers logged into the telco's network. 
They were recorded by order of the nation's domestic intelligence agency.
Further, on her newspaper's website, she allows for a SecureDrop submission for images and videos.
Overall, she \TEAMCASEcircledLetter{TEAMCASEColorWorkflowNumber}{5}receives thousands of hours of audio and video and tens of thousands of texts and images, which she imports into \textsc{MULTI-CASE}.
The system ingests this data and runs the analysis pipeline.

\textbf{Initial Exploration} --- First, Alisa is overwhelmed by the sheer amount of data in the\TEAMCASEcircledLetter{TEAMCASEColorTeaserLabelC}{C}\emph{Knowledge Graph} view.
She looks around and randomly starts listening to some recorded phone calls via the\TEAMCASEcircledLetter{TEAMCASEColorTeaserLabelOne}{1}preview hover menu.
Some are hard to understand due to multiple persons talking intermittently in the background.

\textbf{Analysis Pipeline} --- The system offers her\TEAMCASEcircledLetter{TEAMCASEColorTeaserLabelOne}{1}automatically-generated transcripts through the \emph{Speech to Text} module while the audio is played simultaneously.
She notices that the transcripts are not perfect when hearing the recordings, but they still help her a lot, as she can skim over the content in the \TEAMCASEcircledLetter{TEAMCASEColorTeaserLabelB}{B}\emph{Document Viewer} much quicker.
Wondering if the speakers talked about locations, she searched manually for common city names, finding many results.
She realizes she can also use the entity search to display all locations the semantic text analysis has found, a summary of which is shown at the bottom.
Through these and reading some context, she realizes the transcripts are intermingled with speech fragments (and locations) from the background speakers.

\textbf{Multimodal Combinations} --- She\TEAMCASEcircledLetter{TEAMCASEColorTeaserLabelFive}{5}jumps back to the graph view and selects the \emph{Speaker Recognition} module for the selected node.
It identified four speakers and offered some best shots to listen to, together with individual transcripts.
Hearing them in isolation, she realized that two were radio moderators.
She deselects both speakers and lets the downstream analysis task run again.
In the\TEAMCASEcircledLetter{TEAMCASEColorTeaserLabelC}{C}\emph{Knowledge Graph} view, the old entry is \TEAMCASEcircledLetter{TEAMCASEColorTeaserLabelFour}{4}transparently archived and replaced by the new audio.
Now, the recordings and transcripts are much clearer, but listening or reading through only a few would still take hours.

\textbf{Semantic Search} --- She decides to\TEAMCASEcircledLetter{TEAMCASEColorTeaserLabelA}{A}\emph{search} literally for some terms and words she suspects might have been used but get fewer results.
Instead, she enables the fuzzy as well as the \emph{ontological search}.
Now she receives many more results.
In some, the spelling seems off, and in others, she gets synonyms and hyponyms for her query.
Reading over some of the matching sentences, she realizes several specific words are used and also learns some new ones the system did not detect.

\textbf{Retraining On-The-Fly} --- She adds those words to the \TEAMCASEcircledLetter{TEAMCASEColorTeaserLabelFour}{4}\emph{ built-in ontology} and re-runs the search.
As she reads a conversation about a small village where "a lot of ----- things happened," she feels she might be on to something.
Semantically searching through the remaining transcript in the \TEAMCASEcircledLetter{TEAMCASEColorTeaserLabelB}{B}\emph{Document Viewer}, the speakers refrain from mentioning the village or such events again.

\textbf{Cross-Matches} --- However, the system has recognized the village's name as a location descriptor and offers her to view it in the\TEAMCASEcircledLetter{TEAMCASEColorTeaserLabelC}{C}\emph{Knowledge Graph} view.
There, she uses the\TEAMCASEcircledLetter{TEAMCASEColorTeaserLabelTwo}{2}\emph{Neighborhood Exploration} to see all connected entities up to three steps from this town.
She finds out that another document mentions this tiny village in a spatial context to a larger town while the village is again allegedly mentioned in connection with some persons named $A$ and $B$ repeatedly over an extended time period.
Using the \TEAMCASEcircledLetter{TEAMCASEColorTeaserLabelC}{C}\emph{timeline view}, she restricts the view to a specific time range where she knows that the area around this larger city was temporarily invaded before the attackers were forced out into the neighboring woods.
The graph becomes less crowded, and the system displays a weak link from person $A$ to another name $A`$ with a longer name form.
The weak link comes from yet another transcript, where the persons named are mentioned closely together.

\textbf{Manual Investigations} --- Alisa requests her assistant to read the transcript while she briefs her boss about the preliminary findings.
After returning, Alisa sees that her assistant (working collaboratively on the case with her) has concluded that the persons mentioned in the report are likely similar and\TEAMCASEcircledLetter{TEAMCASEColorTeaserLabelThree}{3} has assigned a B score (highly likely) for the link in the 6x6 system~\cite{UNODC.CriminalIntelligenceManual.2011}.
The person $A`$ has also been mentioned in the caption of a Telegram picture.
Having used the \emph{Image Analyzer} module, her assistant has found visual matches for this person in several pictures and also two videos, which he has flagged for her.
She watches both videos, and one clearly shows a war crime.

\textbf{Handling Fakes} --- She also \TEAMCASEcircledLetter{TEAMCASEColorTeaserLabelA}{A}searches for $B$, and she finds a graphic image but also sees $B$ in a similar setting, seemingly taken weeks prior.
She identifies the environment and obtains a broader view of the situation: the image is fake, likely disinformation.
She \TEAMCASEcircledLetter{TEAMCASEColorTeaserLabelSix}{6}adds a comment and marks it as disproved, becoming archived by default.

\textbf{Progressive Analytics} --- During her background research, interviewing one ICC representative, she is offered access to the ICC evidence collection platform, where users worldwide can upload materials of suspected war crimes.
She also\TEAMCASEcircledLetter{TEAMCASEColorWorkflowNumber}{6}imports this potential evidence enriching the underlying data model.
Now, she runs a further person search using \emph{Image and Video Analyzer} and finds the picture of a military photo ID.
The person in the picture looks very similar to $A$.

\textbf{Evidence Collection} --- Using the\TEAMCASEcircledLetter{TEAMCASEColorWorkflowNumber}{11}reporting functionality, she prints out a trace of her analysis steps, including the transcripts with reference to the original audio files, the connection network with locations, all the associated imagery and data as a PDF report, and the associated document dump.
She plans to hand it over to her ICC contacts and lawyers for them to further verify the potential claims for a subsequent trial.
They plan to perform classical investigative work like forensic audio, facial analysis, and site visitation to collect evidence to back up and corroborate the potential war crime she found using the system, now knowing what to look out for.

\subsection{Ethics Design Guidelines}
\label{sec:teamcase_evaluation_ethics_design}
In previous work~\cite{Fischer.EthicalAwarenessCommAna.2022}, we have discussed in depth the ethical implications of using VA systems in intelligence and derived the first comprehensive overview of \emph{detailed, technical} considerations to take into account when designing such systems.
As pointed out, \emph{the ethical implications [have to be considered] as an integral part of the design process from the outset}~\cite{Fischer.EthicalAwarenessCommAna.2022}.
In the following, we describe how we have applied those considerations during the development of \textsc{MULTI-CASE}.
The guideline numbers (e.g., C1-6, R1-5, A1-6) refer to the nomenclature established in previous work~\cite{Fischer.EthicalAwarenessCommAna.2022}.

Semi-automated analyses are used, but the user remains in control for Human Oversight (R5), and the automated decisions are transparent (e.g., through\TEAMCASEcircledLetter{TEAMCASEColorTeaserLabelFour}{4}attribution and\TEAMCASEcircledLetter{TEAMCASEColorTeaserLabelThree}{3}confidence scoring) for Opacity (C3), addressing User Agency (A1) and Lack of Accountability (R1). \TEAMCASEcircledLetter{TEAMCASEColorWorkflowNumber}{9}Provenance of the analysis steps taken can further strengthen this Human Oversight (R5) and provide Accountability (C6).
The ability, for example, to\TEAMCASEcircledLetter{TEAMCASEColorTeaserLabelSix}{6}flag wrong or unrelated content can support Privacy (A2, C2) aspects by being less intrusive than human verification (as humans might memorize sensitive information).
All automated system risk exhibiting inherent Discriminatory Bias (C1), but human operators also do.
We published our underlying model for transparency reasons (cf. Opacity (C3)) and to detect or Prevent Automated Inequality (R3).
The design as a hybrid Human-Machine-Configurations (C5) inherently \TEAMCASEcircledLetter{TEAMCASEColorTeaserLabelFive}{5} allows for mutual checks and balances to facilitate more Fairness (A3) and Human Oversight (R5).
The semi-automated analysis undoubtedly can \TEAMCASEcircledLetter{TEAMCASEColorTeaserLabelTwo}{2} improve Efficiency (A4), while care was taken not to abstract too much and for the information to remain \TEAMCASEcircledLetter{TEAMCASEColorTeaserLabelOne}{1}transparently attributable (cf. Opacity (C3), Accountability (C6), Lack of Accountability (R1)), which is achieved through the unified \TEAMCASEcircledLetter{TEAMCASEColorWorkflowNumber}{4}fully-integrated data model.
By making clear what aspects are automated and which are manual, by providing\TEAMCASEcircledLetter{TEAMCASEColorTeaserLabelThree}{3}confidence scores, and by not offering unrealistic features such as "solve investigation" buttons, one works against Exaggerated Expectation (C4).
Effective usage of the system and Literacy (A5) can only come with experience and daily usage.
Integrated \TEAMCASEcircledLetter{TEAMCASEColorTeaserLabelSix}{6}sharing between colleagues, e.g., of saved search filters or information through comments, can support this.
However, we note that more could be done here for our approach, but we expect that literacy will primarily be achieved through classical Training and Community-Building Among Users (R2).
By enabling \TEAMCASEcircledLetter{TEAMCASEColorTeaserLabelFour}{4}interactive modifications to the underlying models like the ontologies, Customization (A6) can help users to adapt the system to their needs.
One aspect to further improve upon is automated guidance to facilitate critical reflection (R4), for example, by automatically trying to detect biased behavior by human operators.

\subsection{Intelligence Capability Assessment}
\label{sec:teamcase_evaluation_intelligence_capabilities}
We assess our framework according to a system capabilities classification~\cite{Fischer.CommunicationAnalysis.2022}.
This generic classification aims at knowledge exploration systems, including holistic approaches, focusing on intelligence applications.
The classification's main focus is to assess the (technical) capabilities in a structured form, for example, if time-series data is supported, what type of interactions are used, or which type of knowledge is generated through AI support.
In this regard, it indirectly includes results from older requirements studies in intelligence~\cite{DeckerStasko.VAExploration.2009, Kang.EvaluatingVAInvestigative.2009, Kang.CharacterizingIntelligenceAnalysis.2011, Lu.IntegratingPredictiveAnalytics.2014}.
However, these previous studies primarily describe the user interactions with the system like Jigsaw~\cite{Stasko.Jigsaw.2007} through Overview and Detail, or Find the Clue and Follow the Trial~\cite{Kang.EvaluatingVAInvestigative.2009};
those aspects included in the older studies but not in the capability assessment were evaluated as part of the expert evaluation (see Section~\ref{sec:teamcase_evaluation_expert_study}).
We describe and assess our approach according to the 52 criteria posed in the \textbf{classification scheme}~\cite{Fischer.CommunicationAnalysis.2022}.
The icons indicate \symNo~no, \symPartial~partial, and \symYes~full support.
For a detailed discussion on the attributes themselves, we refer to the original paper while we provide examples and placement of \textsc{MULTI-CASE}'s capabilities in the following:

In the dimension \TEAMCASEhlcolor{TEAMCASEColorBGFrameworkData}{\emph{Data and Information}}, \textsc{MULTI-CASE} can compete with the state-of-the-art:
It supports all basic \textbf{Data types}: \emph{text} like documents and messages, \emph{audio} like recordings, \emph{image/video} like pictures or video recordings, \emph{network} like relationship networks or call records, and \symPartial~\emph{time-series}, primarily through meta-data like discrete timestamps.
Classical, continuous time series are not explicitly supported.
Regarding the \textbf{coding} of data, only \emph{digital} modalities (i.e., the face-value of information) are supported, not \symNo~ \emph{analogical} ones (e.g., interpretation of facial expressions to detect lies or irony).
This is comparable to the vast majority of approaches.
Regarding the orthogonal \textbf{Expression}, \emph{explicit} information is supported, but also \emph{implicit} one, through the use of the underlying ontologies, which is a rare capability.
With regards to communication between \textbf{Parties}, ~\resizebox{!}{.75\baselineskip}{\matrixTT{black}{black}{white}{black}{black}{white}{white}{white}{white}} group communications are supported, but not specifically nested groups (i.e., subgroups).
Analysis of \symNo~\textbf{Power Relations} is not supported.
However, the investigative application is designed in such a way that it accounts for acts of deception and partially considers the \symPartial~ \textbf{Measurement Problem}:
For example, the use of code words is, in principle, supported through the domain-specific ontologies and specially trained NER model and also by looking at meta-data, which is harder to craft.
This is a crucial capability in investigative systems, which many current approaches still delegate fully to the users.

In the dimension \TEAMCASEhlcolor{TEAMCASEColorBGFrameworkModel}{\emph{Processing and Models}}, our approach is suitable for a wide variety of analyses.
Regarding the \textbf{Methodologies}, supported are \emph{Representational} analysis to present the information, and especially \emph{Confirmatory} analysis to validate hypothesis as well as \emph{Exploratory} analysis to find relevant, a priori unknown facts.
\symPartial~\emph{Predictive} Analytics is partly integrated, depending on the employed modules.
In terms of the employed analytical \textbf{Modalities}, all primary ones are equally supported: \emph{Content} like actual text or videos, for example, through the Document Viewer or the Video Analyzer, \emph{Network} for relationship analysis through the Knowledge Graph and Neighborhood Exploration, or \emph{Meta-Data} through the Knowledge Graph and the Timeline in combination with the filtering functions.
This holistic, integrated, and interconnected analysis is a crucial factor distinguishing \textsc{MULTI-CASE} from most existing approaches.
The \textbf{Analysis} itself supports an incremental, streamed data import, making it an \emph{online} analysis.
Regarding the \textbf{Latency}, the standard use case for an investigative system will be a \emph{delayed}~\TEAMCASEsymLetter{D} analysis.
One key advantage of the underlying model and the modular architecture is the achieved \textbf{Scalability}.
It supports huge (IIII) investigative volumes for \emph{ingress}.
Also, through its Neighborhood Exploration, the number of concurrent entries under consideration in the \emph{analysis} can be regarded as medium (II), more than many other approaches.
As our approach is a research prototype and not a commercial application, the support for \symNo~\textbf{Data-Mappings}, like many importers, is limited.

In the dimension \TEAMCASEhlcolor{TEAMCASEColorBGFrameworkVisualization}{\emph{Visual Interface}}, many combined strategies are leveraged.
Regarding the visualization \textbf{Pane}, the usual \emph{2D} is supported, but the Knowledge Graph also leverages \emph{3D}.
Stereoscopic 3D \symNo~\emph{S3D} is unsupported but easily addable.
Regarding the \textbf{Operation Methods}~\cite{Yi.InteractionInfoVis.2007}, all are supported: one can \emph{Select} an entity to get more detailed information from all modules combined, \emph{Explore} different semantic matches or the Knowledge Graph, \emph{Reconfigure} the confidence thresholds for automated merging, \emph{Encode} the data as inferred graph relation representation, \emph{Abstract/Elaborate} by adapting the Neighborhood level or inspect information within a specialized module, \emph{Filter} trough the semantic search or a timeline range, and \emph{Connect} by showing graph neighborhoods.
The \textbf{Manipulation} happens both \emph{directly}, e.g., by selecting entries, and \emph{indirectly}, for example, by choosing specific analysis modes, for example, only showing corroborated cross-matches.
The \textbf{Goal} of the actions is primarily \emph{data tuning} to show relevant information.
However, the approach also partly supports \symPartial~\emph{model tuning}, where the interactions influence an underlying mode, e.g., by manually confirming relationships between entities or updating the ontology.
The \textbf{Strategy} involved in interactions are both \emph{iterative} and \emph{progressive}, which go hand in hand in investigative scenarios.
The \symPartial~\emph{active learning} depends on the individual analysis modules, through feedback or showing an example.

In the dimension \TEAMCASEhlcolor{TEAMCASEColorBGFrameworkKnowledge}{\emph{Knowledge Generation}}, the \textbf{Explanation} of information is performed through \emph{numerical}, \emph{textual}, and \emph{graphical} representations, for example through scoring/sorting, annotated highlighting, or charting, respectively.
The \textbf{Transfer Function} operates both on the \emph{machine model}, which updates the underlying model through interactions, as well as the \emph{mental model} of the analysts.
\textbf{Factors} that are considered in our approach are \emph{confidence}, \emph{trust} and \emph{privacy}.
For a more detailed discussion, see Section~\ref{sec:teamcase_evaluation_ethics_design}.
With regards to the \textbf{Time Dimensionality} of the Knowledge Generation, the approach primarily enables the exploration of past information but also allows conclusions based on this information for the given dataset \matrixDT{black}{black}{white}{white}{white}{white}.
The \textbf{Predictive Power} of the system relates to explaining past events and potential upcoming links but also forms predictions about yet-to-ingest data \matrixDT{black}{black}{black}{black}{white}{white}.
Regarding the \textbf{Evaluations} performed, we present a case study \symTEAMCASEExample (see Section~\ref{sec:teamcase_evaluation_case_study}), this capability assessment \symTEAMCASEComparison (see Sections~\ref{sec:teamcase_evaluation_intelligence_capabilities}-\ref{sec:teamcase_evaluation_ethics_design}), as well as an expert evaluation \symTEAMCASEInterview (see Section~\ref{sec:teamcase_evaluation_expert_study}).

\subsection{Model Evaluation}
\label{sec:teamcase_evaluation_model}
We evaluated \emph{our NER model} (Huggingface model via \OSFLink) and five \emph{baseline models} based on a hold-out test set of the re-tagged news dataset that we publish for future benchmarks.  Based on preliminary experimental results, we decided to train our own NER classifier based on the weights of the pre-trained GottBERT~\cite{Scheible.GottBERT.2020} language model.
We report precision, recall, and F1-score for each entity label (see Tab.~\ref{tab:teamcase_model_ner_accuracy}).
As an overall observation, we find that our test dataset is challenging for the NER model, as the achieved performance is below scores reported for existing benchmark datasets like GermEval2014~\cite{Benikova.Germeval.2014} for all models, including the baselines.

The best-reported score for the GermEval dataset is 86.8\%~\cite{Scheible.GottBERT.2020} with categories \textit{PERSON}, \textit{LOCATION}, \textit{ORGANISATION} and \textit{MISC}.
However, on both our generic news dataset and, in particular, the scenario-specific text data, we see significantly lower performance.
Still, our model outperforms or matches baseline models on the core categories while achieving satisfactory performance on most additional categories. Still, we observe a drop in performance in broad, newly introduced categories like \textit{EVENT} or \textit{PRODUCT}.

\begin{table*}[t]
\centering
\ra{1.1}
\setlength{\tabcolsep}{4.77pt}
\caption[Validation accuracy for the baselines and our model]{%
	\textbf{Validation accuracy} for the five baselines (de\_core\_news\_\{sm,md,lg\}, BERT-GER, XML-RoBERTa) and \emph{our} NER model}
\label{tab:teamcase_model_ner_accuracy}
\begin{tabular}{@{}lccccccccccccccccccccccc@{}}\toprule
	
	& \multicolumn{3}{c}{sm} & \phantom{a}
	& \multicolumn{3}{c}{md} & \phantom{a}
	& \multicolumn{3}{c}{lg} & \phantom{a}
	& \multicolumn{3}{c}{BERT-German} & \phantom{a}
	& \multicolumn{3}{c}{XML-RoBERTa} & \phantom{a}
	& \multicolumn{3}{c}{\textbf{Ours}}\\
	\cmidrule{2-4} \cmidrule{6-8} \cmidrule{10-12} \cmidrule{14-16} \cmidrule{18-20} \cmidrule{22-24}
	Type
	& P & R & F1 
	& & P & R & F1 
	& & P & R & F1 
	& & P & R & F1 
	& & P & R & F1 
	& & P & R & F1  \\

	\midrule
	PERSON
	& .69 & .72 & .70
	& & .76 & .77 & .77
	& & .78 & .78 & .78
	& & .93 & \textbf{.89} & \textbf{.91}
	& & .91 & .87 & .89
	& & \textbf{.94} & .88 & \textbf{.91} \\
	
	ORGANIZATION
	& .55 & .47 & .51
	& & .56 & .52 & .54
	& & .59 & .55 & .57
	& & .81 & .65 & .72
	& & .75 & .64 & .69
	& & \textbf{.78} & \textbf{.78} & \textbf{.78} \\
	
	LOCATION
	& .53 & .57 & .54
	& & .59 & .61 & .60
	& & .61 & .61 & .61
	& & \textbf{.90} & .63 & .74
	& & .84 & .62 & .71
	& & .88 & \textbf{.90} & \textbf{.89} \\

	MISC (Original)
	& .14 & .29 & .19
	& & .17 & .37 & .24
	& & .18 & .36 & .24
	& & - & - & -
	& & \textbf{.30} & \textbf{.45} & \textbf{ .36}
	& & - & - & - \\
	
	MISC (Own)
	& - & - & - & - & - & - & - & - & - & - & - & - & - & - & - & - & - & - & - & - &
	.15 & .22 & .18 \\
	
	EVENT
	& - & - & - & - & - & - & - & - & - & - & - & - & - & - & - & - & - & - & - & - &
	.99 & .40 & .57 \\
	
	PRODUCT
	& - & - & - & - & - & - & - & - & - & - & - & - & - & - & - & - & - & - & - & - &
	.49 & .59 & .54 \\
	
	DATETIME
	& - & - & - & - & - & - & - & - & - & - & - & - & - & - & - & - & - & - & - & - &
	.99 & .99 & .99 \\
	
	LANGUAGE
	& - & - & - & - & - & - & - & - & - & - & - & - & - & - & - & - & - & - & - & - &
	.98 & .95 & .96 \\
	
	LAW
	& - & - & - & - & - & - & - & - & - & - & - & - & - & - & - & - & - & - & - & - &
	.60 & .60 & .60 \\
	
	QUANTITY
	& - & - & - & - & - & - & - & - & - & - & - & - & - & - & - & - & - & - & - & - &
	.97 & .96 & .97 \\
	
	NUMBERS
	& - & - & - & - & - & - & - & - & - & - & - & - & - & - & - & - & - & - & - & - &
	.98 & .98 & .98 \\
	
	\bottomrule
\end{tabular}
\end{table*}

\subsection{Domain Expert Evaluation}
\label{sec:teamcase_evaluation_expert_study}
To showcase the effectiveness of our approach in comparison to existing methods, we conducted an expert evaluation with eleven domain experts (\DELEA{1-2}, \DERS{1-3}, \DESI{1-3}, \DELAW{1}, \DEEE{1}, \DEPOL{1}) working in the context of law enforcement.

\textbf{Expertise}
\DELEA{1} is a recently retired former special police forces commander with a 40-year career, now working as a security consultant for law enforcement agencies.
\DELEA{2} is a leading investigator at a federal police force with a 20-year career investigating organized crime.
\DERS{1} is a research scientist and head of the research department with a 30-year career in speech recognition.
\DERS{2} is a junior researcher and developer working for a federal security agency developing analytical solutions for law enforcement in the area of digital forensics.
\DERS{3} is a junior researcher working for the same federal security agency on the topic of phone analysis.
\DESI{1} is a senior principal research engineer with an almost 30-year career overseeing numerous identity solution projects for an international security company.
\DESI{2} is a project manager with a 25-year career working on video analysis and investigative systems at the same company.
\DESI{3} is a principal research scientist with more than ten years of experience in video object tracking also at this company.
\DELAW{1} is a professor and criminologist specializing in security management, hate crimes, and legal aspects with more than 15 years of experience in the field.
\DEEE{1} is a sociologist and ethics advisor offering guidance for security research projects.
\DEPOL{1} is a project supervisor at a national project management agency overseeing civil security research and policy expert.

\textbf{Methodology} The expert evaluation was conducted as a formative evaluation and took a combined 180 minutes, split into a 60-minute presentation and a 120-minute hands-on evaluation.
The 60-minute introduction delivered to all experts described the capabilities of the system on a conceptual level while also demonstrating actions in the form of one to three-minute-long screen recordings.
During the evaluation, a single station (27-inch FHD screen, mouse, and keyboard) with the prototype was available to the experts, together with two researchers standing by to help with questions and advice.
During this time, one of the experts would typically use the system to explore the prototype while being encouraged to think aloud.
The other experts could meanwhile observe, comment, and ask questions.
After irregular intervals, the experts switched positions, and usage time between experts varied between five to 20 minutes.
During the whole session, the experts were asked questions aligned with a semi-structured interview sheet containing a set of 38 prepared questions covering various aspects of our approach.
The session's aim was to elicit the domain experts' opinions about the system and gain insights into how they would use the system in their investigative workflows.
Further, the experts were asked to comment on the approaches' capabilities, user-interaction concepts, and visualizations while identifying opportunities for improvements.
The detailed findings of this evaluation are presented in the following.

\textbf{Findings} 
Asked about the \textbf{benefits} they see in an investigative framework like \textsc{MULTI-CASE}, the criminal investigators state that they hoped to be relieved of the time-consuming, \enquote{extremely high manual workload, which currently requires much personnel} (\DELEA{1}) \enquote{and time} (\DELEA{2}).
Of course, there are existing use case management systems, but \enquote{their usage and the casework is performed very much in a manual way [$\ldots$] with little technical support} (\DELEA{1}), which becomes a \enquote{big problem for mass data} (\DELEA{2}), where \enquote{automation can be very helpful} (\DERS{2}).
In \enquote{particular observations produce very large amounts of video data} (\DELEA{1}).
For particular problems, some isolated technical solutions exist at some local partners, for example, geo-based analyses (cf. \DELEA{1}), but access depends on the local support and willingness of the partners to help (cf. \DELEA{1}).
Further, one of the most important features for them is to import many different types of multimodal documents like \enquote{existing records, images, videos} (\DELEA{2}). 
Here, \textsc{MULTI-CASE} as a \TEAMCASEcircledLetter{TEAMCASEColorTeaserLabelOne}{1} \enquote{large overview system for multimodal data like audio, text or video has the potential to drastically improve investigative work} (\DERS{3}), making it \enquote{uncharted territory} (\DELEA{1}).
The other experts strongly agree, noting that currently they lack \enquote{a complete picture [in a single system]} (\DELEA{1}) and \enquote{nothing in this form exists} (\DELEA{1}): neither for phones (cf. \DERS{3}), speech (cf. \DELEA{2}), or text (cf. \DELEA{1}).
\enquote{Multimodality is the largest benefit, as everything can be seen in context} (\DERS{2}).

Regarding the \textbf{risk of automation}, they are aware of potential pitfalls but do not consider them highly problematic:
It is likely that "there are errors in the analysis" (\DERS{1}), for example, by different spellings (cf. \DERS{1}).
This, however, can also happen when case workers need \enquote{to read through thousands of pages or watch weeks of video recordings, where things might be overlooked and error rates increase with time as frustration increases} (cf. \DELEA{1}).
\enquote{From an automated perspective, it might not be most important to find everything, but to start and find many relevant things} (\DEPOL{1}).
From a \enquote{legal perspective, this might be much more critical, as innocent individuals can become part of an investigation} (\DERS{1}).
They note that \enquote{automated analysis is less of a problem when there is reasonable suspicion for a suspect, but an infringement on fundamental rights is} (\DELAW{1}).
In this regard, the modality differs: \enquote{images are considered more critical than voice, which in turn is more critical than text} (cf. \DELAW{1}).
Possible ways to solve this are \enquote{by not focusing on the subject, but on the right infringements [for involved parties]} (cf. \DELAW{1}).
This means automated analysis has the potential to be considered less invasive than manual analysis, but \enquote{for example through data economy and short-term storage, but this depends on the case} (\DEEE{1}).

From an \textbf{ethical perspective}, it might be more \enquote{justifiable to let the computer search for targets instead of humans} (\DEEE{1}) as the human \enquote{remembers} (\DEEE{1}) offering potential for misuse, while the system forgets after the comparison.
Current approaches \enquote{do not consider privacy or ethical aspects sufficiently} (\DELEA{2}) and the investigators are independently responsible on their own to follow the rules - however, \enquote{there is a wide gap between theory and practice} (\DELEA{2}).
\enquote{A verified system that works with high accuracy [and without bias] could be fairer than an arbitrary human} (\DEEE{1}), as \enquote{many humans are very selective and inherently biased} (\DELEA{1}).
Regarding the fear of intransparent, autonomous decisions, it was noted that "the systems are always support systems and humans always the final instance" (\DELEA{1}), and before a \enquote{prosecution will always be manually verified} (\DELEA{1}).
A problem can arise when \enquote{humans become too careless and trust the system too much} (\DEEE{1}).

One interesting discussion arose regarding the error rate:
From the perspective of an analyst \enquote{false positives are less of an issue as they can be manually verified, while false negatives are missed} (\DELEA{1}).
From the \enquote{perspective of innocents, this is directly inverse, but this again depends on the context} (cf. \DEEE{1}). \enquote{When misses lead to extreme dangers for others, this can be very bad} (cf. \DEEE{1}). 

The experts consider\TEAMCASEcircledLetter{TEAMCASEColorTeaserLabelSix}{6}\textbf{collaboration} features relevant, where multiple users can work on the same case, as they sometimes have to work with \enquote{widely distributed experts} (cf. \DELEA{1}).
Also, the \enquote{parallel work between colleagues is nice} (\DEEE{1}).

Regarding the central \textbf{Knowledge Graph}, many experts agree that it can provide a key overview, as \enquote{it is extremely important to show all the relations} (\DEPOL{1}) and the \enquote{connections} (\DELEA{1}), which is a \enquote{large advantage} (\DERS{2}). 
\enquote{Showing everything together is very relevant for keeping an overview} (\DESI{1}).
For this, the\TEAMCASEcircledLetter{TEAMCASEColorTeaserLabelTwo}{2}\emph{Neighborhood Exploration} is considered \enquote{a must, especially when many data items are loaded} (\DERS{2}), as it allows to reduce the visual clutter and only show contextual information.
This is an example of a filtering functionality, which is regarded as \enquote{essential} (\DERS{2}).
Also, the ability to filter the graph and the mergings by\TEAMCASEcircledLetter{TEAMCASEColorTeaserLabelThree}{3}confidence is regarded to be beneficial (\DERS{1}).
Similarly, the timeline is also considered \enquote{very helpful} (\DERS{2}), as \enquote{the time and event sequence is very important for the investigation} (\DELEA{2}).
In this context, the interactions are regarded as \enquote{very smooth and nice looking} (\DERS{2}).
However, some experts questioned \enquote{if 3D is necessary} (cf. \DELEA{1}) and would favor the 2D graph that is also available.
The graph view can act as a \enquote{supportive mental map [$\ldots$] and a large digital notebook} (\DELEA{2}), which \enquote{currently is often only in ones head} (\DELEA{2}).
For this, the \enquote{comment function} is essential and helpful (cf. \DELEA{2} and \DEEE{1}) to make notes, which can be shared between users.
Regarding the confirmatory investigative work, however, the \enquote{graph view is less important} (\DELEA{1}), where the \enquote{individual analysis modules like the document viewer or audio analysis are more relevant [$\ldots$] supporting the daily work} (\DELEA{1}).
For example, in the document viewer, the \enquote{automated recognition of entities in the document which are shown at the bottom with their number of occurrences, is especially helpful, as it allows to get a\TEAMCASEcircledLetter{TEAMCASEColorTeaserLabelFour}{4}summary understanding of the content of the text already} (cf. \DELEA{1}).
Also, the automated transcription of audio \enquote{given sufficient quality, is very important and a key advantage} (cf. \DELEA{1}).
Especially relevant is the ability to seamlessly switch between view and modalities, for example, \TEAMCASEcircledLetter{TEAMCASEColorTeaserLabelFive}{5}\enquote{to jump from a node in the graph to the text location in the document viewer} (\DELEA{1}) as well as \enquote{jumping to search matches}  (\DELEA{1}). However, it was noted by several experts that a proficient usage would \enquote{require training} (cf. \DELEA{1}, \DEEE{1}, \DERS{1}, \DELAW{1}), after its completion, however, would be a \enquote{productivity boost} (\DEEE{1}).

In terms of \textbf{potential future features}, some ideas were mentioned:
Among expected quality-of-life improvements like more file type support (cf. \DERS{2}),
one area of improvement could be group conversations (cf. \DERS{2}), for example, through colored attributions also inside the document viewer, the creation of cluster-nodes in the graph view to merge related, but currently less interesting entities (cf. \DERS{2}) or show a modification and usage history from co-workers (cf. \DERS{3}).
Also, for the comments and exploration history of colleagues, a \enquote{misuse button} (cf. \DEEE{1}) would potentially be useful to report incorrect use.
Also, some more explainability for the automated parts, i.e., why a \enquote{speaker was recognized} (cf. \DEEE{1}) as such, would be useful and increase trust.
Overall, the approach \enquote{will be well usable for semi-automated investigative analysis [$\ldots$] between a knowledgeable user and a supportive system} (\DELEA{1}).

\section{Discussion and Future Work}
\label{sec:teamcase_discussion}
As we demonstrated, our approach enhances the capabilities for multimodal intelligence analytics.
In the following, we discuss the valuable lessons we learned during development, the implications of the valuable feedback we received about our prototype, architectural design trade-offs, limitations of the approach, and potential future work that remains.

\subsection{Findings and Lessons Learned}
\label{sec:teamcase_findings}
Based on the evaluations in the previous section, we have succeeded in working towards fulfilling the experts' requirements posed from the beginning:
\textsc{MULTI-CASE} is an exemplary centralized, multimodal platform framework that allows several analysts to collaboratively work on cases and empowers users through the transparent inclusion of AI-aided decision-making while relieving them of burdensome tasks and considering ethical design guidelines.
Following the UNODC~\cite{UNODC.CriminalIntelligenceManual.2011} task definitions, the main tasks can be performed:
link analysis between entities is supported while also allowing to consider them in the context of the surrounding events based on a timeline.
While it supports a basic flow analysis in principle, the visualization modules presented here are not particularly suited for this analysis, but through the modular design, a component operating on the shared data model could be developed.
We have seen how the multimodal approach can support the analysis of otherwise difficult-to-detect cross-matches, while a visual analytics-based approach has benefits in terms of agency, accountability, and trust.
The experts are open to AI-based solutions, especially when it relieves them of mundane tasks, and they feel supported.
Leveraging both computational power and human intuition in a tight feedback loop can positively influence the capabilities of the resulting human-machine configuration.
Regarding the displayed results, they tend to believe them at face value to some degree when they seem plausible, somewhat similar to findings reported to them by colleagues.

However, we also saw that experts have high expectations regarding the machine results and---especially when not specifically trained for the system---are rather unforgiving with respect to unexpected or contradictory results.
Also, they can be easily annoyed in case they feel the system hinders them, holds them back, or torments them through seemingly obvious confirmations.
Based on these observations, we can derive several key findings:

\begin{leftbar}{TEAMCASEColorInfoBoxTitle}{TEAMCASEColorInfoBoxBody!50}
\noindent\textbf{F1: A Holistic Approach Supports Finding Cross-Matches} \newline
The case study and expert evaluation shows that intelligence investigations require interconnected, multimodal analytics.\newline
\textit{Implication:}  A holistic approach can combine these different analysis modalities within a single context, reducing domain boundaries and enabling effective search for cross-matches.
Especially relevant here is a vertical integration between all analysis modules, for example, through a fully-integrated data model.%
\end{leftbar}

\begin{leftbar}{TEAMCASEColorInfoBoxTitle}{TEAMCASEColorInfoBoxBody!50}
\noindent\textbf{F2: Unobtrusive Support-Systems are Accepted} \newline
As long as a system remains supportive and unobtrusive, relieving analysts of mundane tasks or providing them with valuable hints and insights on request or through nudging, semi-automated systems are accepted. Tormenting approaches hindering the workflow or being intuitive or unreliable can destroy an initial level of trust placed in the system.  \newline
\textit{Implication:} A self-explanatory, easy-to-use user interface combined with helpful but unobtrusive functions is essential. For this, the right balance has to be found between automation and manual confirmation. Unreliable or inconsistent results (without indications) or hindering of workflows should be strongly avoided %
\end{leftbar}

\begin{leftbar}{TEAMCASEColorInfoBoxTitle}{TEAMCASEColorInfoBoxBody!50}
\noindent\textbf{F3: Reduce False Negatives for VA---and False Positives for AI} \newline
Initially surprising to us was that for many tasks (e.g., search, filter, linking), the domain experts (both \DELEA{1-2} and \DEEE{1}) prefer the error rate to depend on the automation level: for semi-interactive VA a reduction of false-negatives is often preferable, while automated systems should reduce false-positives.\newline
\textit{Implication:} Consider the optimization task carefully, as the \emph{cost of error}, where not finding something (i.e., FN) or a wrong attribution (i.e., FP) is considered more costly than the opposite.
A missed lead might break the whole investigation, while a wrong attribution might cause serious harm to innocents.%
\end{leftbar}

\begin{leftbar}{TEAMCASEColorInfoBoxTitle}{TEAMCASEColorInfoBoxBody!50}
\noindent\textbf{F4: Limited Acceptance of Unreasoned Decisions} \newline
At least for now, to support an ethical and privacy-aware analysis and offer transparency, fairness, and accountability while fostering user trust, the experts prefer an explainable, interactive system compared to a fully automated approach. \newline
\textit{Implication:} Due to the high stakes in this domain, experts have concerns about fully-automated systems that cannot provide a rigorous chain of evidence, which---at least for now---is rarely possible.
Future developments might shift this balance. %
\end{leftbar}
\vspace*{0.1cm}

\subsection{Limitations and Future Work}
\label{sec:teamcase_limitations_future_work}
Nevertheless, the approach remains a research project with limitations:

The \textbf{Knowledge Graph} representation uses custom GPU-optimized rendering achieving excellent performance, but it comes with the disadvantage that some of the more advanced results from graph drawing, like more complex curved lines, are not directly applicable without heavy performance penalties.
We also want to highlight that we do not see our contribution in designing state-of-the-art graph drawing but in the interactions, combinations, and linkings between the different modalities for the graph.

The integration of the \textbf{underlying language model} itself is modular, such that any other transformer-based NER model can be easily used, as the system features a built-in language detection.
However, for the evaluation in this chapter, we only used our customized German NER model due to the domain experts' preferences and expertise.
We did not explicitly show a generalization, which we, nevertheless, certainly expect.
In the future, off-the-shelf transformer-based NER models can be used, with limitations in the types of detected NER and resulting degradation in relationship inference.
Alternative models would need to be fine-tuned with additional NER types, requiring appropriate training data.
Another problem in this regard can be the analysis of multi-lingual or inter-lingual text and transcripts.

The recent progress with \textbf{Large Language Models (LLMs)} like GPT-4~\cite{OpenAI.GPT.2023} offers interesting opportunities in this regard.
This is, in particular, relevant when models are capable of supporting multiple languages as well as providing up-to-date and case-specific query context, as the \emph{New Bings} underlying Prometheus Model~\cite{Microsoft.NewBing.2023} shows to some limited degree.
Three domain experts (\DELEA{1}, \DESI{1}, \DESI{2}) in our study tried Chat-GPT on crafted case material and were astonished both by the easy workflow of querying and the (relative) quality of the findings as potential leads.
They regard such \emph{text-based, interactive prompting} through LLMs, which imitates basic reasoning and summarization capabilities, as potentially very useful.
Integrating such natural language prompts in applications, maybe only in supportive roles, seems very promising.
Interestingly, GPT-4 also shows surprising capabilities in zero-shot NER labeling.
For testing, we let GPT-4 auto-label a subset of our test data. We achieved this zero-shot labeling by prepending a prompt "Extract named entities of the given types from the following text: person, organization and location", resulting in only slightly less quality than manual, human labeling.
This could potentially replace specifically trained NER models, like the one we described in Section~\ref{sec:teamcase_methodology}.
Recent experiments~\cite{Gilardi.ChatGPTAnnotation.2023} suggest superior results are possible.
While this shows the viability of the transfer learning approach, "close-to-real-life" scenarios often perform worse compared to controlled benchmarks~\cite{Paleyes.ChallengesDeployML.2022}.
Therefore, evaluating such scenarios in the wild is important to identify persisting limitations, which can be supported by interactive analysis.
Also, care must be taken to consider the additional risks involved when using LLMs: They do not learn from mistakes outside their limited context window (32k for GPT-4-32k), which is relevant when using all documents as context, and most seriously, they tend to suffer from hallucinations that are hard to detect.
Further, employment of such solutions would require on-premise solutions or specialized contracts.

Overall, depending on the jurisdictions, \textbf{legal requirements} might regulate the allowed automated analysis tasks~\cite{BVerfG.Palatir.2023}.
The ethical and privacy-aware design, as well as the semi-automated analysis, always subject to human verification, performed in our approach, should allow for usage even in tightly regulated jurisdictions.
The concrete usage in critical cases, however, should be accompanied by a prior legal counsel.

One general limitation in this line of research is the \textbf{opaqueness of the intelligence community}.
Many systems are classified~\cite{Dhami.SurveyIntelligenceAnalystToolsPerception.2017} and capabilities are not shared openly--which can be frustrating from a scientific perspective, hampering progress and introducing problems from ethical and privacy perspectives due to missing accountability.
It also remains difficult to recruit domain experts to evaluate and analyze the techniques developed in the scientific community.
One way to reduce increased reliance on expert evaluations is to also incorporate general interaction strategy design guidelines derived from numerous user interaction evaluations regarding relevance feedback~\cite{Khan.InteractionStrategiesUserRelevance.2021}.
Efforts are ongoing to finance more research in open and accountable intelligence solutions (e.g., within Horizon Europe and others).
However, we are well aware that some aspects of this domain will likely remain hidden.
With our work, we try to contribute to ongoing research in this domain and discuss ways to make these more accountable.

\section{Conclusion}
\label{sec:teamcase_conclusion}
Over the last few years, AI-driven models have become increasingly prevalent in many domains.
This tendency can also be observed in operational analytics solutions in investigative journalism, intelligence, or law enforcement.
These domains, in particular, pose distinct challenges due to their sensitive nature.
Two aspects, in particular, stand out: ethical and privacy concerns, as well as difficulties in efficiently combining heterogeneous data sources for multimodal analytics.
A lack of such \emph{holistic and multimodal} approaches can lead to biased results and increased manual efforts through domain discontinuities.

To address these two challenges, we present \textsc{MULTI-CASE}, a holistic visual analytics framework that enables the exploration and assessment of heterogeneous information spaces (i.e., unstructured, diverse, and multimodal) supported by an equal joint agency between humans and AI to ensure ethics and privacy awareness.
To fulfill these requirements, the system operates on a fully-integrated data model while featuring type-specific analyses with multiple linked components, including a modality-wide search (i.e., full-text, semantics, and all multimodal analysis results), text, and graph-based analysis.
Different information streams are linked in a knowledge graph, providing in-situ explanations and transparent source attributions while facilitating responsible exploration through numerous interlinked explorative modules.
We discuss the potential for improvements, for example, in rendering, completeness, or the use of more advanced LLMs.

We demonstrate how our framework fulfills the design goals through state-of-the-art intelligence capability assessments and evaluations according to ethics design guidelines.
The underlying transformer model showed state-of-the-art performance on relevant benchmarks.
To showcase our prototype's analytical capabilities in practice, we presented a case study describing war crime investigations in the context of investigative journalism.
Finally, a formative expert evaluation with eleven domain experts in law enforcement confirms that \textsc{MULTI-CASE} facilitates human agency and steering in security-sensitive, AI-supported analysis, addresses ethical and privacy concerns, and provides much-needed analytical capabilities.

With this contribution, we aim to provide more insights into the often opaque workings of the intelligence community and strive towards a more accountable and responsible use of modern AI capabilities.

\section*{Acknowledgments}
The authors acknowledge the financial support by the Federal Ministry of Education and Research of Germany (BMBF) in the framework of PEGASUS and VIKING under the program "Forschung für die zivile Sicherheit 2018 - 2023" and its announcement "Zivile Sicherheit - Schutz vor organisierter Kriminalität II".

\bibliographystyle{IEEEtran}
\bibliography{bibliography}

\vfill

\end{document}